\documentclass[pdflatex,sn-mathphys-num]{sn-jnl}

\usepackage{graphicx}%
\usepackage{multirow}%
\usepackage{amsmath,amssymb,amsfonts}%
\usepackage{amsthm}%
\usepackage{mathrsfs}%
\usepackage[title]{appendix}%
\usepackage{xcolor}%
\usepackage{textcomp}%
\usepackage{manyfoot}%
\usepackage{booktabs}%
\usepackage{algorithm}%
\usepackage{algorithmicx}%
\usepackage{algpseudocode}%
\usepackage{listings}%

\usepackage{graphicx}
\usepackage{subcaption}

\usepackage{etoolbox}
\apptocmd{\thebibliography}{\setlength{\itemsep}{1pt}}{}{}

\theoremstyle{thmstyleone}%
\theoremstyle{thmstyletwo}%

\theoremstyle{thmstylethree}%

\begin{document}

\title[Investigating Algorithmic Bias in YouTube Shorts]{Investigating Algorithmic Bias in YouTube Shorts} 


\author[1]{\fnm{Mert Can} \sur{Cakmak}}\email{mccakmak@ualr.edu}

\author[1,2]{\fnm{Nitin} \sur{Agarwal}}\email{nxagarwal@ualr.edu}

\author[1]{\fnm{Diwash} \sur{Poudel}}\email{dpoudel@ualr.edu}



\affil[1]{COSMOS Research Center, University of Arkansas, Little Rock, USA}

\affil[2]{ICSI, University of California, Berkeley, USA}




\abstract{The rapid growth of YouTube Shorts, now serving over 2 billion monthly users, reflects a global shift toward short-form video as a dominant mode of online content consumption. This study investigates algorithmic bias in YouTube Shorts’ recommendation system by analyzing how watch-time duration, topic sensitivity, and engagement metrics influence content visibility and drift. We focus on three content domains: the South China Sea dispute, the 2024 Taiwan presidential election, and general YouTube Shorts content. Using generative AI models, we classified 685,842 videos across relevance, topic category, and emotional tone. Our results reveal a consistent drift away from politically sensitive content toward entertainment-focused videos. Emotion analysis shows a systematic preference for joyful or neutral content, while engagement patterns indicate that highly viewed and liked videos are disproportionately promoted, reinforcing popularity bias. This work provides the first comprehensive analysis of algorithmic drift in YouTube Shorts based on textual content, emotional tone, topic categorization, and varying watch-time conditions. These findings offer new insights into how algorithmic design shapes content exposure, with implications for platform transparency and information diversity.}

\keywords{YouTube Shorts, Algorithmic Bias, Recommendation Systems, Generative AI, Content Drift, Watch-Time, Political Content, Popularity Bias}



\maketitle

\section{Introduction}

Short-form video platforms such as YouTube Shorts, TikTok, and Instagram Reels have reshaped how users engage with digital content. These platforms rely heavily on recommendation algorithms that prioritize engagement, often leading to biases in content visibility and emotional tone. YouTube Shorts, launched in 2021, now attracts over 2 billion monthly logged-in users, highlighting its growing influence on public discourse \cite{rajendran2024shorts}. Given the scale of YouTube's influence, with over 30\% of Americans regularly using the platform as a news source \cite{StAubin2024}, understanding how its algorithms shape content visibility is both timely and necessary. Research has shown that YouTube Shorts tends to favor viral or emotionally engaging content, while deprioritizing complex or serious topics such as politics and education \cite{violot2024shorts}. This creates an algorithmic bias that can narrow the diversity of recommended content, limit user exposure to critical issues, and shape public perception in subtle ways. To explore this phenomenon, we examine three distinct content domains: the South China Sea (SCS) dispute, the 2024 Taiwan presidential election, and general YouTube Shorts content. The SCS dispute remains a critical geopolitical flashpoint involving territorial claims, economic interests, and military tensions between China, the Philippines, the U.S., and other regional actors \cite{seo2024power}. The Taiwan election was a major political event in early 2024, raising concerns over disinformation, foreign influence, and democratic resilience \cite{NYTimesTaiwan2024}. These politically sensitive topics offer a compelling contrast to the more diverse and entertainment-driven general Shorts content. By including this broader category, we provide a baseline for comparison and improve the generalizability of our findings across different content domains. This study investigates the potential biases in YouTube Shorts recommendations by addressing the following research questions:

\begin{itemize}
    \item \textbf{RQ1:} To what extent does content drift occur and how does it impact the relevance and diversity of suggested videos?
    \item \textbf{RQ2:} What types of content are disproportionately promoted or suppressed?
    \item \textbf{RQ3:} How does watch-time influence the bias in recommendation algorithm?
\end{itemize}

To answer these questions, we collect and analyze Shorts data using generative AI models to classify video content by relevance, topic, and emotion. We also incorporate engagement metrics and simulate different watch-time conditions to investigate how these factors influence algorithmic behavior. By comparing politically sensitive and general topics, we uncover systematic patterns of drift and favoritism that shape the Shorts recommendation ecosystem. To the best of our knowledge, this is the first study to comprehensively analyze algorithmic behavior on YouTube Shorts using textual features, emotional tone, topic dynamics, and engagement under varying watch-time settings. Since YouTube’s recommendation logic is proprietary, our analysis relies on observed outputs from controlled and simulated interactions. This is a common approach in algorithm auditing. As short-form content continues to dominate online video consumption, addressing algorithmic bias is essential to ensuring that platforms like YouTube promote information diversity alongside entertainment. This study contributes to that goal by offering a detailed analysis of how algorithmic decisions affect visibility and engagement, particularly for underrepresented or serious topics.


\section{Literature Review} \label{Literature-Section}

Short-form video platforms such as YouTube Shorts, TikTok, and Instagram Reels have rapidly transformed content consumption and user engagement. These platforms rely on recommendation algorithms that prioritize engagement metrics, which can introduce biases affecting content visibility, diversity, and emotional tone. A growing body of research has explored these biases, particularly in the contexts of content drift, filter bubbles, and the influence of algorithmic design on user experience. In the case of YouTube, several studies have shown that the platform's algorithms tend to favor popular and highly engaging content. This behavior can reinforce content homogeneity and marginalize niche or minority topics \cite{kirdemir_2021_advances, cakmak2025influence}. Content drift, where recommendations gradually move away from a user's original interests, has also been widely observed. For example, \cite{cakmak_2024_analyzing} found that topical relevance in recommendations declines over time, particularly in multilingual and cross-regional contexts. Emotional drift is similarly prevalent. Studies such as, \cite{cakmak_2024_investigating}, and \cite{cakmak2024bias} reported that content with joyful or neutral sentiment is promoted over content expressing anger, fear, or other negative emotions, especially when dealing with politically sensitive topics.

These biases tend to be even more pronounced in short-form video platforms. \cite{yang2022bias} demonstrated that TikTok's ``For You" page prioritizes viral content, creating a cycle where popular creators and themes are continuously amplified. \cite{cakmak2024unveiling} identified biases in thumbnail recommendations within YouTube Shorts, where visually appealing or emotionally charged thumbnails were more likely to be recommended. Related research by \cite{hunnego2024exploring} has shown how personalization and filter bubbles in platforms like Instagram Reels and TikTok can reinforce existing beliefs and limit exposure to diverse viewpoints. Watch-time has also emerged as a key factor in shaping recommendation patterns. \cite{zhang_2023_leveraging} emphasized that watch-time provides more nuanced signals of user engagement compared to clicks or likes. In short-form video platforms, this metric plays a central role in retaining viewer attention, which may further encourage the algorithm to promote emotionally engaging and entertaining content at the expense of serious or minority-focused topics.

Despite this growing body of research, YouTube Shorts remains understudied. Most existing work has focused on TikTok and Instagram Reels, often emphasizing visual content or user behavior. To the best of our knowledge, no prior research has systematically examined recommendation drift in YouTube Shorts using textual features, topic content, watch-time conditions, and emotion analysis. This study addresses this gap by introducing a scalable generative AI-based framework to analyze algorithmic behavior across multiple dimensions. By investigating how politically sensitive and general content is treated in terms of relevance, topic, and emotional tone, we offer new insights into content drift and engagement bias. Our findings contribute to ongoing discussions about algorithmic transparency, platform accountability, and equitable content exposure in short-form video environments.

\section{Methodology}\label{Methodology}

This section outlines our data collection process, the application of generative AI models for large-scale content classification, and the validation of these models for accurate relevance, topic, and emotion labeling.

\subsection{Data Collection}

To investigate algorithmic bias and content drift on YouTube Shorts, we designed a structured data collection process across three thematic areas: (1) the 2024 Taiwan presidential election, (2) the South China Sea territorial disputes, and (3) a broad set of general YouTube content categories. This approach allows us to examine how YouTube’s recommendation system behaves across politically sensitive and general domains. The Taiwan election and South China Sea disputes were selected due to their geopolitical sensitivity, where risks of information manipulation are well-documented. By analyzing them independently, we assess whether YouTube’s algorithm responds differently to serious political topics. In contrast, collecting a broader, topic-agnostic dataset enables us to study algorithmic behavior within mainstream and trending content.

\textbf{Keyword Collection} — For the \textit{Taiwan election dataset}, we curated keywords related to electoral events, disinformation narratives, candidate-specific campaigns, and international relations, based on journalistic and academic sources \cite{klepper_wu_2024, NYTimesTaiwan2024}. The \textit{South China Sea (SCS) dataset} was developed through expert workshops with researchers from the Atlantic Council’s Digital Forensic Research Lab, the National University of Singapore, and De La Salle University, focusing on regional disputes, military developments, and environmental impacts \cite{atlanticcouncil_dfrlab, nus_oceanlaw_southchinasea}. For broader coverage, we compiled general content categories following YouTube’s taxonomy \cite{entreresource_youtube_categories_2023}, including entertainment, education, news, sports, and science. Table~\ref{tab:keywords} summarizes the keywords used for data collection.

\begin{table}[h]
\caption{Summary of Keywords Used for Data Collection. For the Taiwan Election and South China Sea datasets, representative keywords are shown. For the General Content dataset, the full set of categories is included.}
\label{tab:keywords}
\begin{tabular*}{\textwidth}{@{\extracolsep\fill}lp{0.8\textwidth}}
\toprule
\textbf{Dataset} & \textbf{Keywords (Selected)} \\
\midrule
Taiwan Election & Taiwan 2024 election, Lai Ching-te wins Taiwan election, Taiwan election disinformation, China interference Taiwan election, Taiwan election cybersecurity, Taiwan election misinformation, Taiwan election social media influence. \\
South China Sea & South China Sea dispute, Scarborough Shoal standoff, US Navy South China Sea, China maritime claims, South China Sea latest news, China vs Philippines, South China Sea oil and gas, ASEAN South China Sea talks. \\
General Content & Film \& Animation, Autos \& Vehicles, Music, Pets \& Animals, Sports, Travel \& Events, Gaming, People \& Blogs, Comedy, Entertainment, News \& Politics, How-to \& Style, Education, Science \& Technology, Nonprofits \& Activism. \\
\botrule
\end{tabular*}
\end{table}

\textbf{YouTube Shorts Collection} — Since the official YouTube Data API v3 does not support the collection of Shorts content, we used APIFY, a web scraping platform, to gather YouTube Shorts video IDs. APIFY simulates organic user browsing by loading Shorts based on keyword queries, similar to how results appear on YouTube. This approach reflects typical user interaction and does not introduce additional selection bias beyond the keyword design. Specifically, we utilized the YouTube Scraper tool \cite{streamers_2024_youtube}, configuring it to collect only Shorts by setting the \texttt{maxResultsShorts} parameter to a large value while disabling other video types (e.g., live, regular videos). For the Taiwan election dataset, we filtered Shorts published between January 9 and January 25, 2024 to improve relevance, covering the pre- and post-election period (the election was held on January 13, 2024). No date filters were applied to the South China Sea or general content datasets to capture a broader temporal range. The final dataset sizes were 2,137 videos for general content, 2,124 for South China Sea, and 281 for Taiwan election. The smaller Taiwan dataset was expected due to the narrower collection window.

\textbf{YouTube Shorts Recommendation Collection} — To investigate the recommendation dynamics of YouTube Shorts, we developed a custom scraping framework, as no existing methodology supports direct collection of Shorts recommendations. This novelty adds value to our study. Using the previously collected video IDs as seeds, we initiated recommendation sessions. Each seed video was opened in a fresh WebDriver instance with no browsing history, no cookies, and no logged-in user, ensuring a neutral environment. Automation was implemented using the Selenium Python library, simulating user interaction by scrolling through recommended Shorts to a depth of 50, where depth is defined as the position in the recommendation chain (e.g., depth 1 is the first recommendation). After each session, the browser was fully closed and relaunched to prevent session contamination. A depth of 50 was chosen to balance data comprehensiveness with practical runtime constraints, aligned with typical user engagement patterns. To study watch-time effects, we collected recommendations under three viewing conditions: short watch (3 seconds), moderate watch (15 seconds), and full watch (up to 60 seconds), representing varying levels of user engagement. Manual scrolling delays were incorporated to better simulate natural user behavior. Across all datasets and watch-time settings, we collected a total of exactly 685,842 videos, including both root and recommended videos, as summarized in Table~\ref{tab:recommendation_stats}. For each video, we retrieved metadata such as title, view count, like count, and comment count using the YouTube Data API v3. Descriptions were excluded due to their frequent absence in Shorts. Transcripts were extracted by following established methodologies \cite{cakmak_2023_adopting, cakmak_2024_high}. This  data collection framework enables a robust analysis of YouTube Shorts' recommendation dynamics across different topical domains and user engagement levels.

\begin{table}[ht]
\centering
\caption{Summary of Root and Recommended Videos Collected. Recommended videos were collected under three different watch-time settings (3s, 15s, full watch), resulting in total counts approximately three times larger than single-setting collections.}
\begin{tabular}{ccc}
\hline
\textbf{Dataset} & \textbf{Root Videos} & \textbf{Recommended Videos (Single Setting) (Total)} \\
\hline
General Content & 2,137 & 106,850 (320,550) \\
South China Sea & 2,124 & 106,200 (318,600) \\
Taiwan Election & 281 & 14,050 (42,150) \\
\hline
\end{tabular}
\label{tab:recommendation_stats}
\end{table}

\subsection{Algorithmic Drift}

To examine algorithmic drift on YouTube Shorts, we analyzed changes in \textit{relevance}, \textit{topic}, and \textit{emotional tone} across recommendation chains. Our aim was to measure whether the content recommended by YouTube's algorithm remained consistent with the initial seed video in terms of its relevance and thematic or affective properties, or whether it gradually diverged over time. We first assessed \textbf{relevance drift} by comparing the semantic relevance of each recommended video to the original topic. The goal was to evaluate whether recommendations maintained topical alignment or shifted toward unrelated content as depth increased. Next, we analyzed \textbf{topic drift}. Given our focus on political content and existing literature showing that YouTube Shorts often prioritizes entertainment, we categorized videos into three groups: \textit{politics}, \textit{non-entertainment}, and \textit{entertainment}. This classification enabled us to observe whether content transitioned from serious to lighter themes across the recommendation depth. We also examined \textbf{emotional drift}, motivated by the tendency of entertainment content to express joy or humor, while political content often conveys neutral or negative affect. Emotion was measured from the video’s title and transcript, based on expressed tone rather than viewer response. We define emotion as expressed in the video’s textual content (title and transcript), not the emotion elicited in viewers. While these can differ, our goal is to detect systematic shifts in the tone of recommended content, rather than subjective audience response. Finally, we evaluated \textbf{engagement levels}—including views, likes, and comments—to investigate whether the algorithm disproportionately promotes high-engagement content, potentially reinforcing popularity bias. To enable scalable and consistent labeling, we employed Generative AI models to classify relevance, topic, and emotion for each video. These models offer high efficiency, objectivity, and reproducibility for subjective classification tasks. Recent studies have demonstrated the utility of large language models in content moderation and affective analysis \cite{jooesten_2024_comparing}.

\subsection{Validation of Generative AI Models for Classification Tasks}\label{secA1}

For all classification tasks (relevance, topic, emotion), we used the video title and transcript as input. We evaluated three generative AI models: GPT-4o, Meta-LLaMA-3, and Gemini 1.5 Pro, to classify relevance, topic, and emotion. Validation was performed using six benchmark datasets across these tasks, with prompts and datasets summarized in Table~\ref{tab:datasets_prompts_compact}. As shown in Table~\ref{tab:genai_model_scores}, GPT-4o consistently outperformed the other models. Consequently, GPT-4o was used for all downstream algorithmic drift analyses.


\begin{table}[h]
\caption{Benchmark Datasets and Prompt Instructions Used for Model Validation. Prompts were designed for deterministic label generation. In the relevance prompt, \textit{X} refers to either the \textbf{South China Sea dispute} or the \textbf{Taiwan election}.}
\label{tab:datasets_prompts_compact}
\begin{tabular*}{\textwidth}{@{\extracolsep\fill}l p{0.4\textwidth} p{0.42\textwidth}}
\toprule
\textbf{Category} & \textbf{Dataset Description} & \textbf{Prompt Instruction} \\
\midrule
Emotion & 
\textbf{GoEmotions}~\cite{dorottya_goemotions}: 58k Reddit comments labeled with 27 fine-grained emotions or neutral; 
\textbf{DailyDialog}~\cite{yanran_dailydialog}: Multi-turn conversations labeled with emotion distributions. & 
\footnotesize\textit{You are a model that classifies a given sentence according to emotion categories. Please provide only the assigned emotion label, as follows: Print ‘0’ for happiness/joy, ‘1’ for sadness, ‘2’ for anger, ‘3’ for neutral, and ‘4’ for fear.} \\
\midrule
Topic & 
\textbf{BBC News}~\cite{jikadara_bbc_news}: News articles from BBC between 2022 and 2024 categorized by topic; 
\textbf{News Category}~\cite{misra_news}: HuffPost headlines collected from 2012 to 2022, labeled into 41 categories. & 
\footnotesize\textit{You are a model that classifies a given sentence according to topic categories. Please provide only the assigned topic label, as follows: Print ‘0’ for politics, ‘1’ for non-entertainment, and ‘2’ for entertainment topics.} \\
\midrule
Relevance & 
\textbf{MS MARCO}~\cite{bajaj_marco}: Human-annotated passages with relevance scores based on real web search queries; 
\textbf{WikiQA}~\cite{yang_wikiqa}: Question-answer sentence pairs from Wikipedia for open-domain QA. & 
\footnotesize\textit{You are a model that classifies a given sentence according to its relevance to the X topic. Please provide only the assigned relevance label as follows: Print ‘3’ for high relevance, ‘2’ for moderate relevance, ‘1’ for low relevance, and ‘0’ for no relevance.} \\
\botrule
\end{tabular*}
\end{table}

\begin{table}[h]
\centering
\caption{Performance Comparison of Generative AI Models Across Emotion, Topic, and Relevancy Tasks}
\label{tab:genai_model_scores}
\begin{tabular}{lccc}
\hline
\textbf{Dataset / Category} & \textbf{LLaMA 3 (\%)} & \textbf{GPT-4o (\%)} & \textbf{Gemini 1.5 (\%)} \\
\hline
\multicolumn{4}{|c|}{\textbf{Emotion Classification}} \\
\hline
GoEmotions & 46.7 & 61.5 & 38.7 \\
DailyDialog & 52.8 & 72.4 & 50.6 \\
\textbf{Average (Emotion)} & 49.75 & \textbf{66.95} & 44.65 \\
\hline
\multicolumn{4}{|c|}{\textbf{Topic Classification}} \\
\hline
BBC News & 70.41 & 88.2 & 35.9 \\
News Category & 62.84 & 82.1 & 33.7 \\
\textbf{Average (Topic)} & 66.63 & \textbf{85.15} & 34.8 \\
\hline
\multicolumn{4}{|c|}{\textbf{Relevancy Matching}} \\
\hline
MS MARCO & 71.80 & 80.46 & 69.84 \\
WikiQA & 77.50 & 77.00 & 80.60 \\
\textbf{Average (Relevancy)} & 74.65 & \textbf{78.73} & 75.22 \\
\hline
\end{tabular}
\end{table}

Since no labeled datasets exist for YouTube Shorts or similar platforms, we used standard benchmark datasets for validation. While not domain-specific, they offer a practical proxy given the lack of short-form video annotation resources.

\section{Results} \label{Results}

This section presents the results of our analysis across four key dimensions: relevance, topic distribution, emotional tone, and user engagement. Each analysis examines how content evolves across recommendation depth and watch-time settings, with particular attention to differences between politically sensitive topics and general content.

\subsection{Relevance Levels}

\begin{figure}[htbp]
  \centering

  \begin{subfigure}[b]{0.32\textwidth}
    \includegraphics[width=\linewidth]{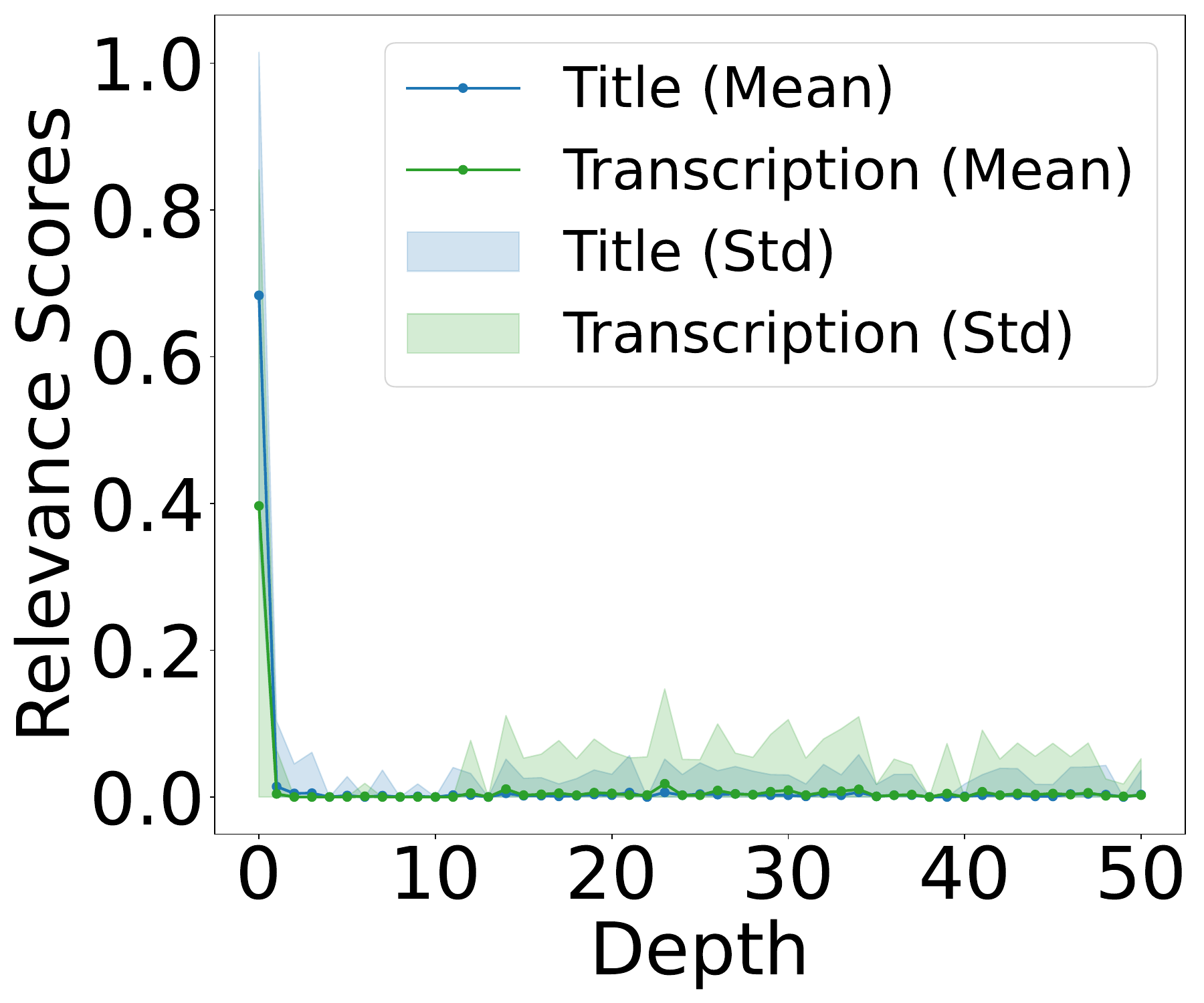}
    \caption{\centering \shortstack{SCS - Relevance Levels\\(3 Secs)}}
    \label{scs-rel_3}
  \end{subfigure}
  \hfill
  \begin{subfigure}[b]{0.32\textwidth}
    \includegraphics[width=\linewidth]{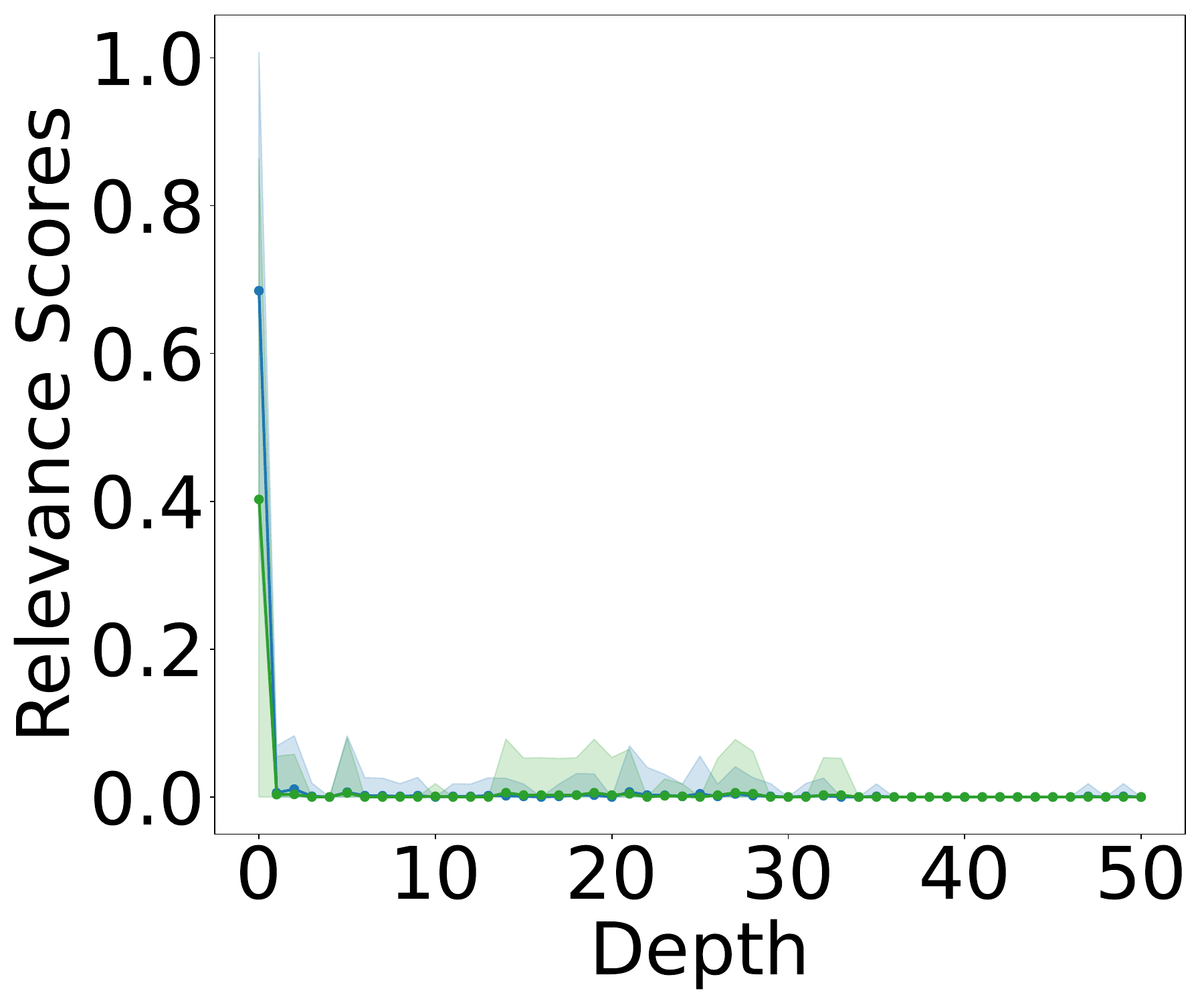}
    \caption{\centering \shortstack{SCS - Relevance Levels\\(15 Secs)}}
    \label{scs-rel_15}
  \end{subfigure}
  \hfill
  \begin{subfigure}[b]{0.32\textwidth}
    \includegraphics[width=\linewidth]{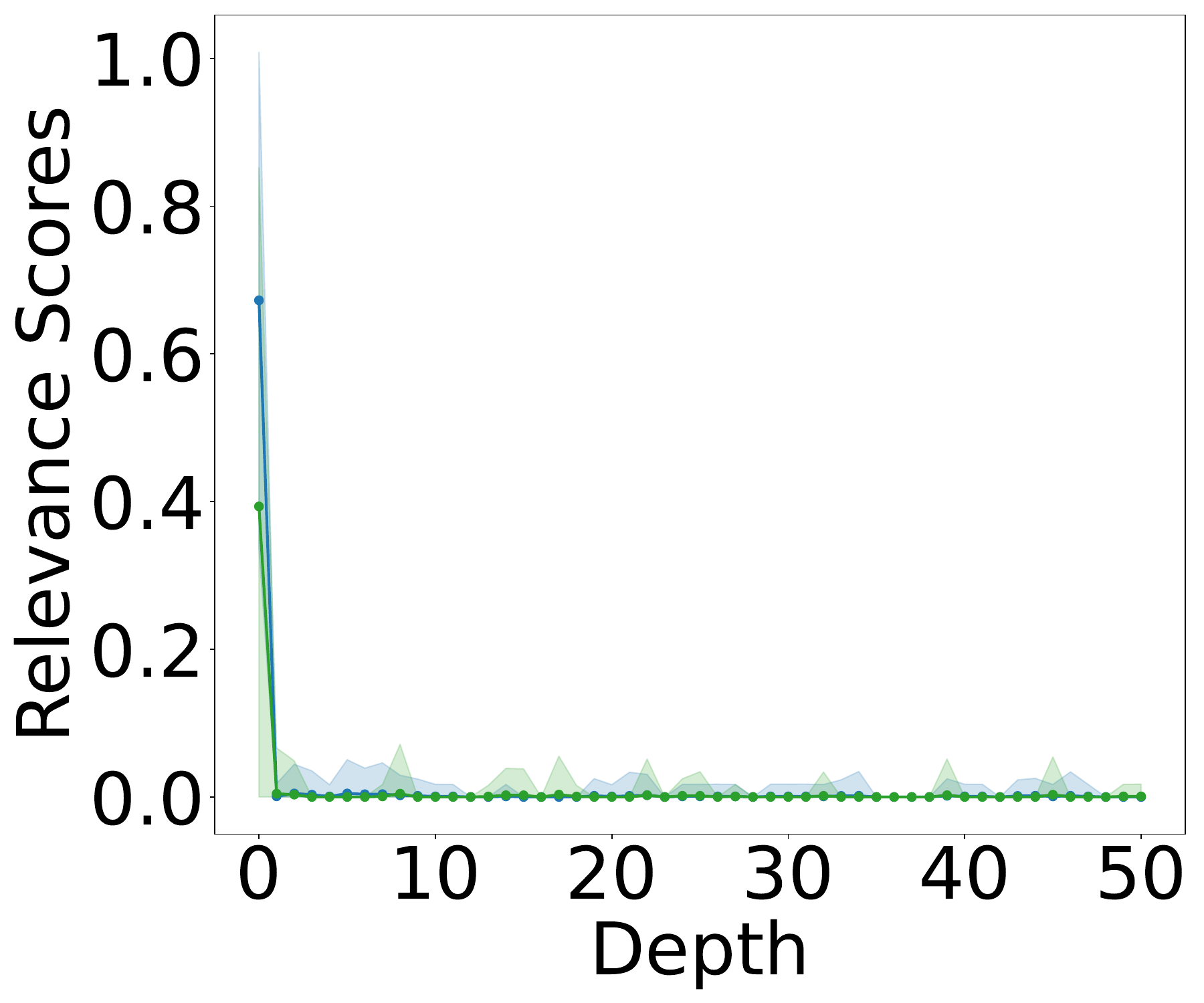}
    \caption{\centering \shortstack{SCS - Relevance Levels\\(60 Secs)}}
    \label{scs-rel_60}
  \end{subfigure}

  \vspace{0cm}

  \begin{subfigure}[b]{0.32\textwidth}
    \includegraphics[width=\linewidth]{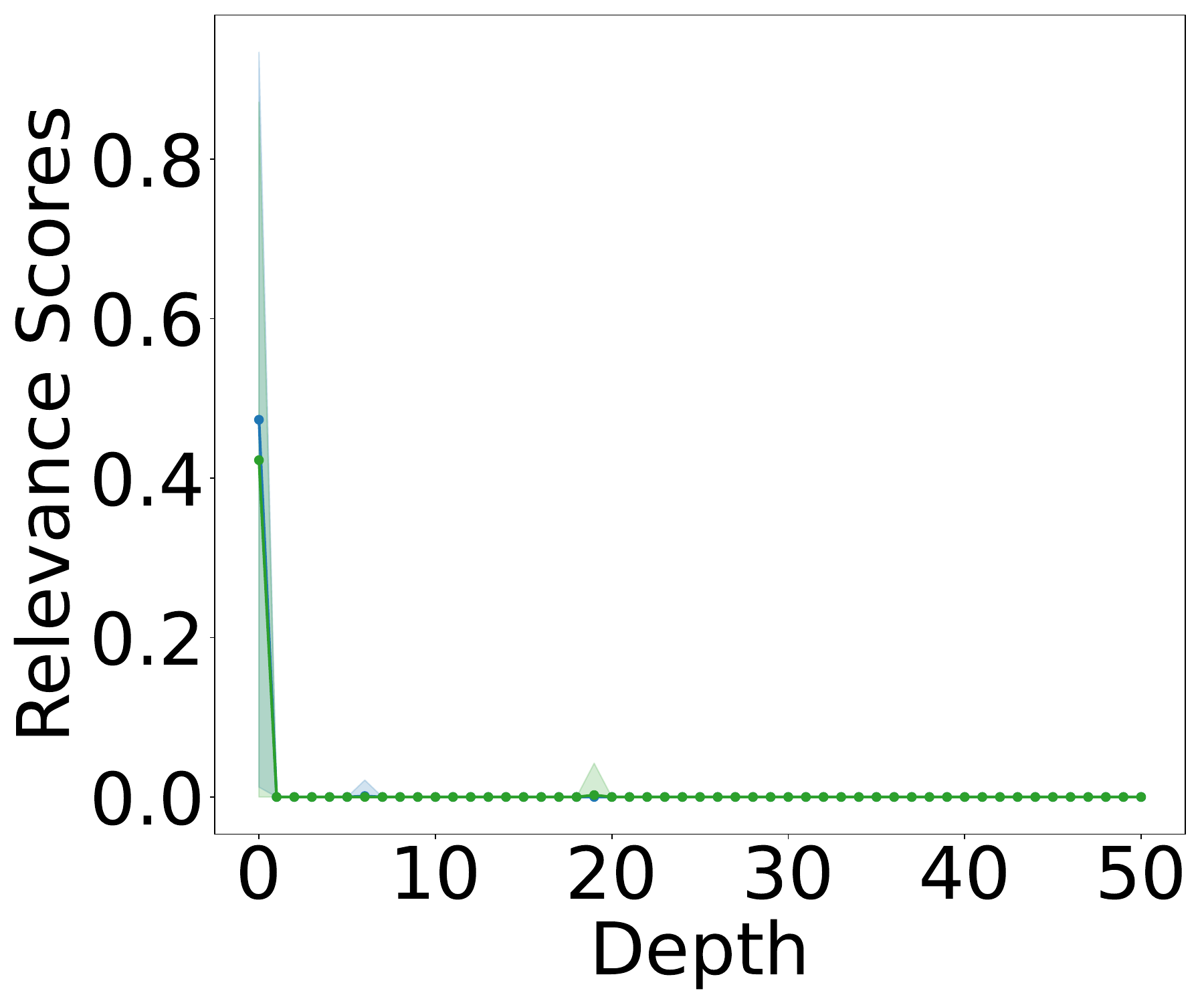}
    \caption{\centering \shortstack{Taiwan - Relevance Levels\\(3 Secs)}}
    \label{taiwan-rel_3}
  \end{subfigure}
  \hfill
  \begin{subfigure}[b]{0.32\textwidth}
    \includegraphics[width=\linewidth]{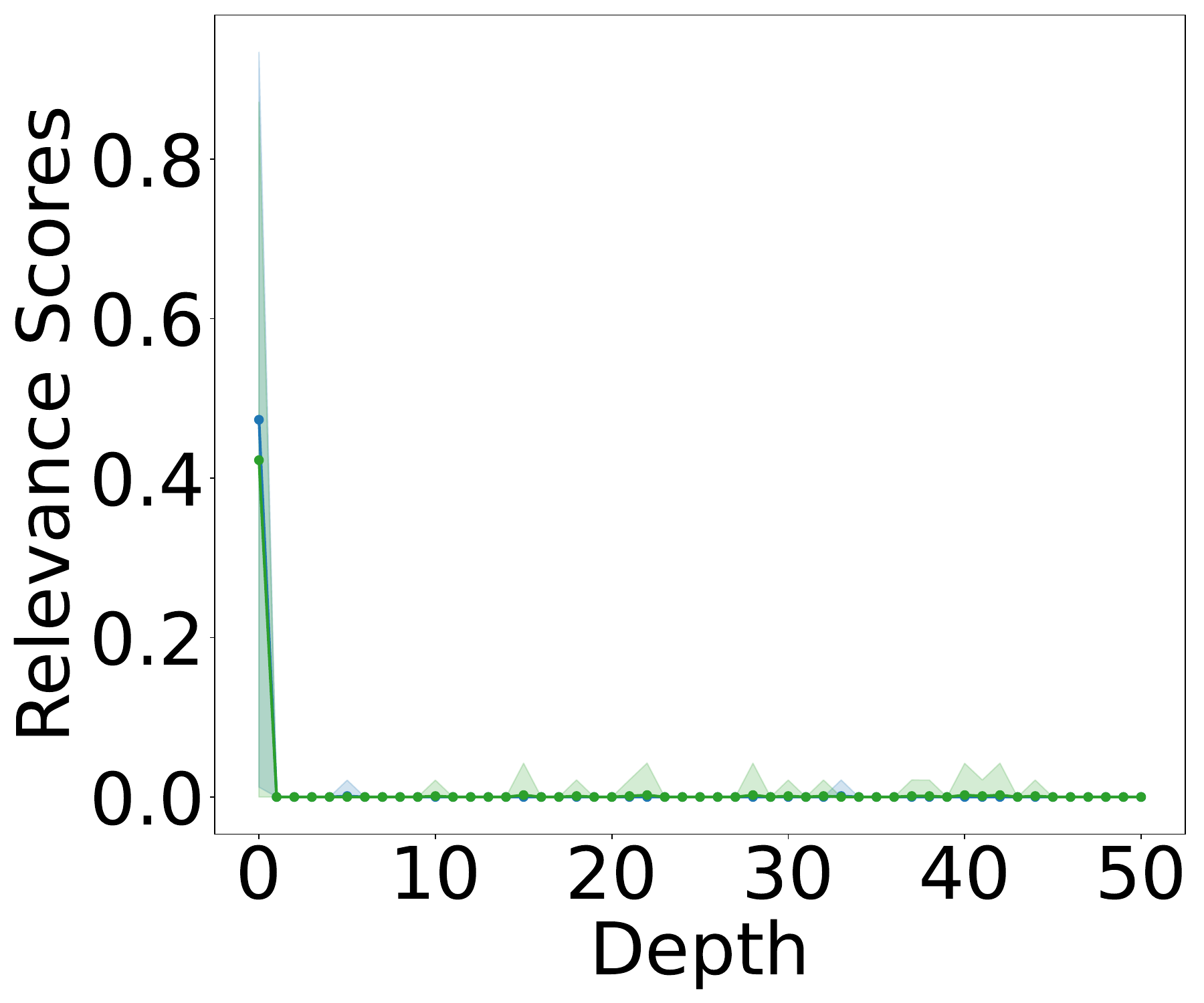}
    \caption{\centering \shortstack{Taiwan - Relevance Levels\\(15 Secs)}}
    \label{taiwan-rel_15}
  \end{subfigure}
  \hfill
  \begin{subfigure}[b]{0.32\textwidth}
    \includegraphics[width=\linewidth]{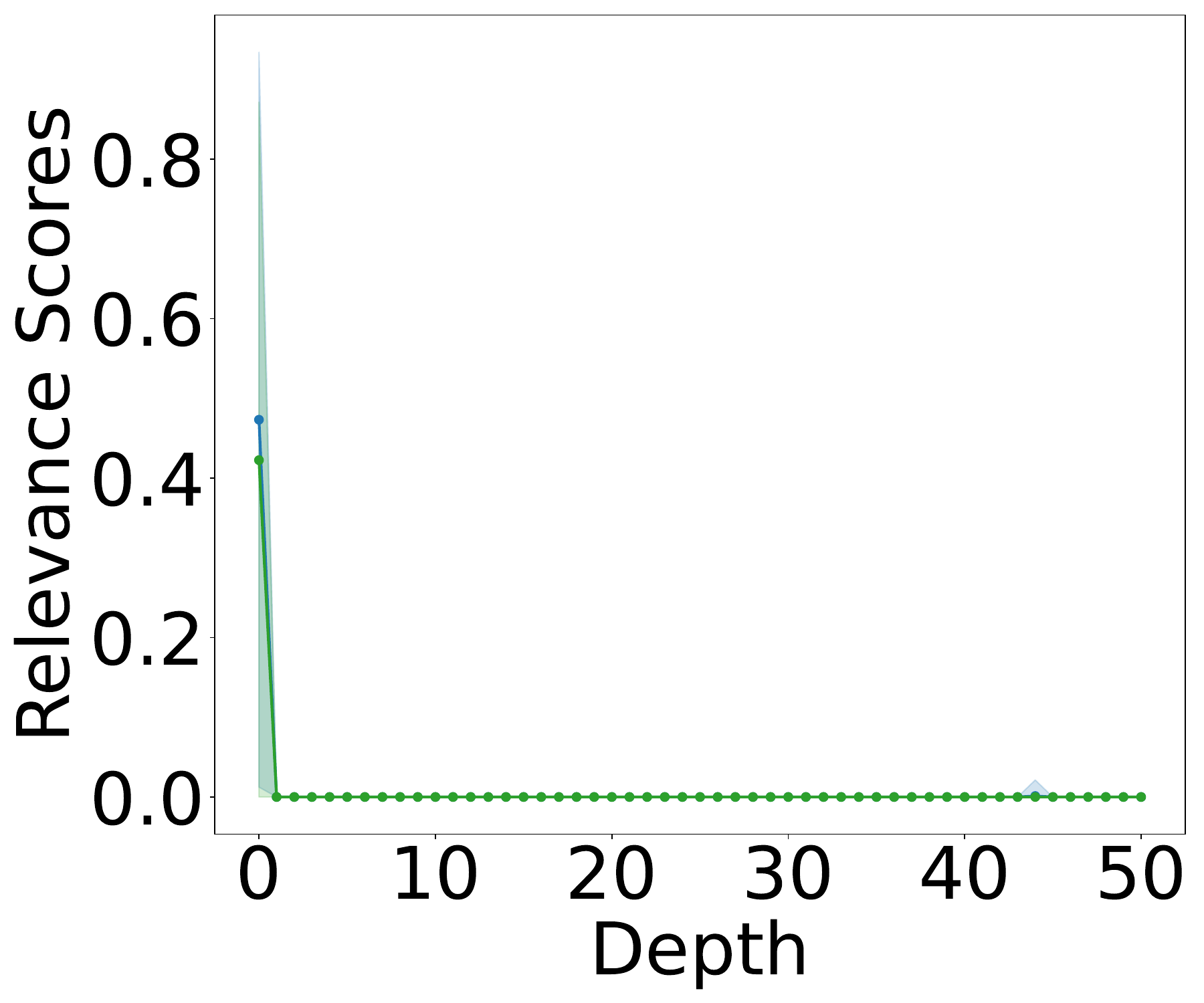}
    \caption{\centering \shortstack{Taiwan - Relevance Levels\\(60 Secs)}}
    \label{taiwan-rel_60}
  \end{subfigure}

  \caption{Relevance scores to the initial topic at each recommendation depth for the South China Sea (SCS) and Taiwan Election datasets. The lines show the mean values and standard deviations.}
  \label{fig:all_relevance}
\end{figure}

To assess how well YouTube Shorts maintains topical consistency during recommendations, we conducted a relevance-based analysis. Recommended videos were scored for semantic proximity to the initial topic, categorized into four levels: high, medium, low, and none. Scores were normalized from a 0–3 scale to a 0–1 range. This analysis focused on the South China Sea and Taiwan Election datasets, which represent well-defined topics. We excluded the General Content dataset due to its diverse, topic-agnostic nature, which lacks a clear thematic anchor. As shown in Figure~\ref{fig:all_relevance}, both narratives exhibited high or medium relevance at depth 0, reflecting strong alignment in seed videos. Titles generally scored higher than transcriptions, as they were more concise and topically focused. In contrast, transcriptions often included off-topic or filler content, resulting in lower relevance. However, relevance dropped sharply at depth 1 across all watch-time conditions (3s, 15s, and full duration), approaching near-zero values. This suggests that the algorithm diverges from the original topic almost immediately, regardless of user engagement. These results indicate that watch-time duration does not meaningfully impact topic relevance for politically sensitive content. The algorithm deprioritizes the original theme shortly after the initial video, even under full engagement. In the following sections, we extend this analysis to topic and emotion distributions, incorporating results from the General Content dataset to provide a broader view of recommendation behavior.

\subsection{Topic Distribution}
\begin{figure}[!ht]
  \centering

  \begin{subfigure}[b]{0.32\textwidth}
    \includegraphics[width=\linewidth]{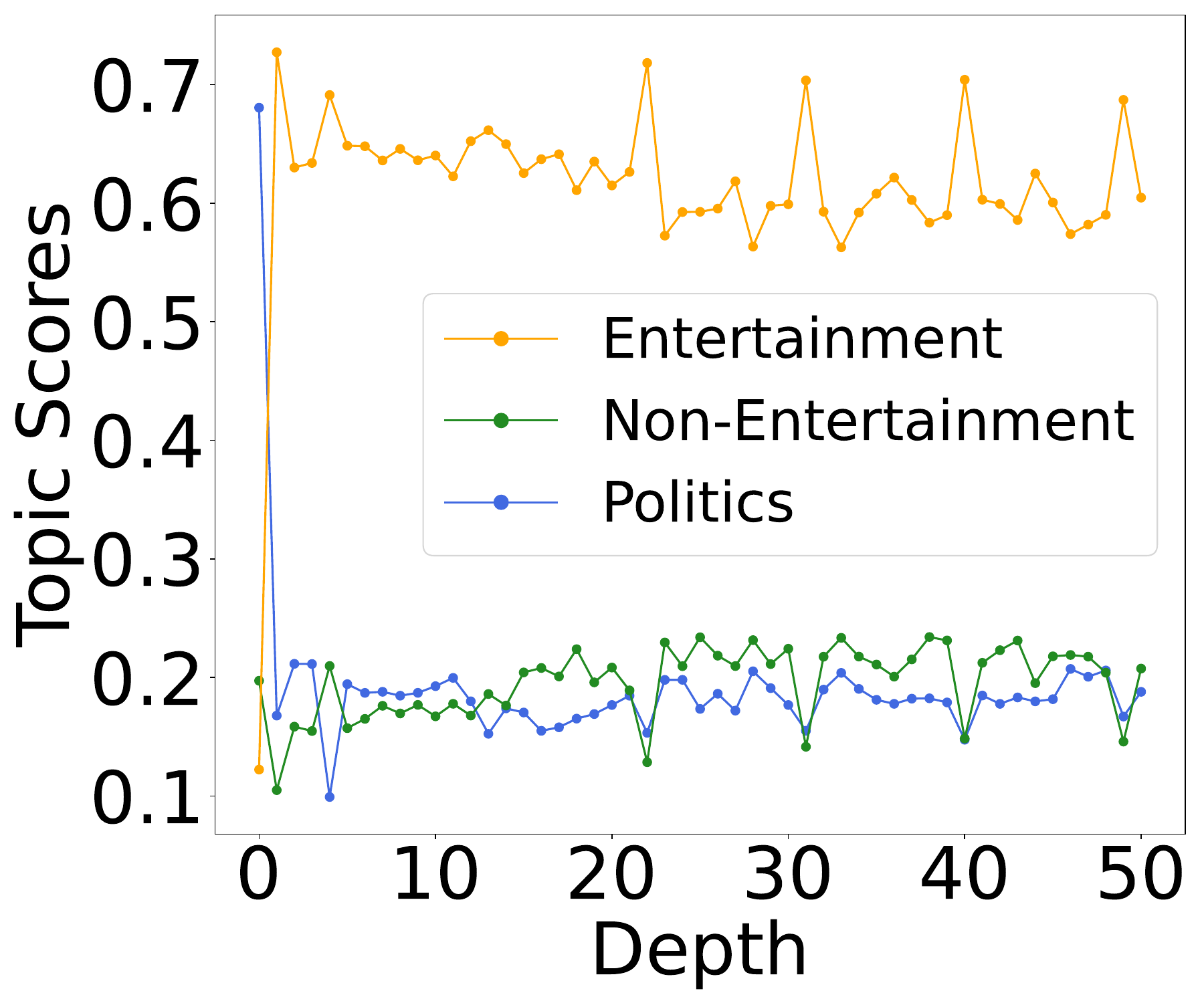}
    \caption{\centering \shortstack{SCS Topic Levels\\(3 Secs)}}
    \label{scs-topic_3}
  \end{subfigure}
  \hfill
  \begin{subfigure}[b]{0.32\textwidth}
    \includegraphics[width=\linewidth]{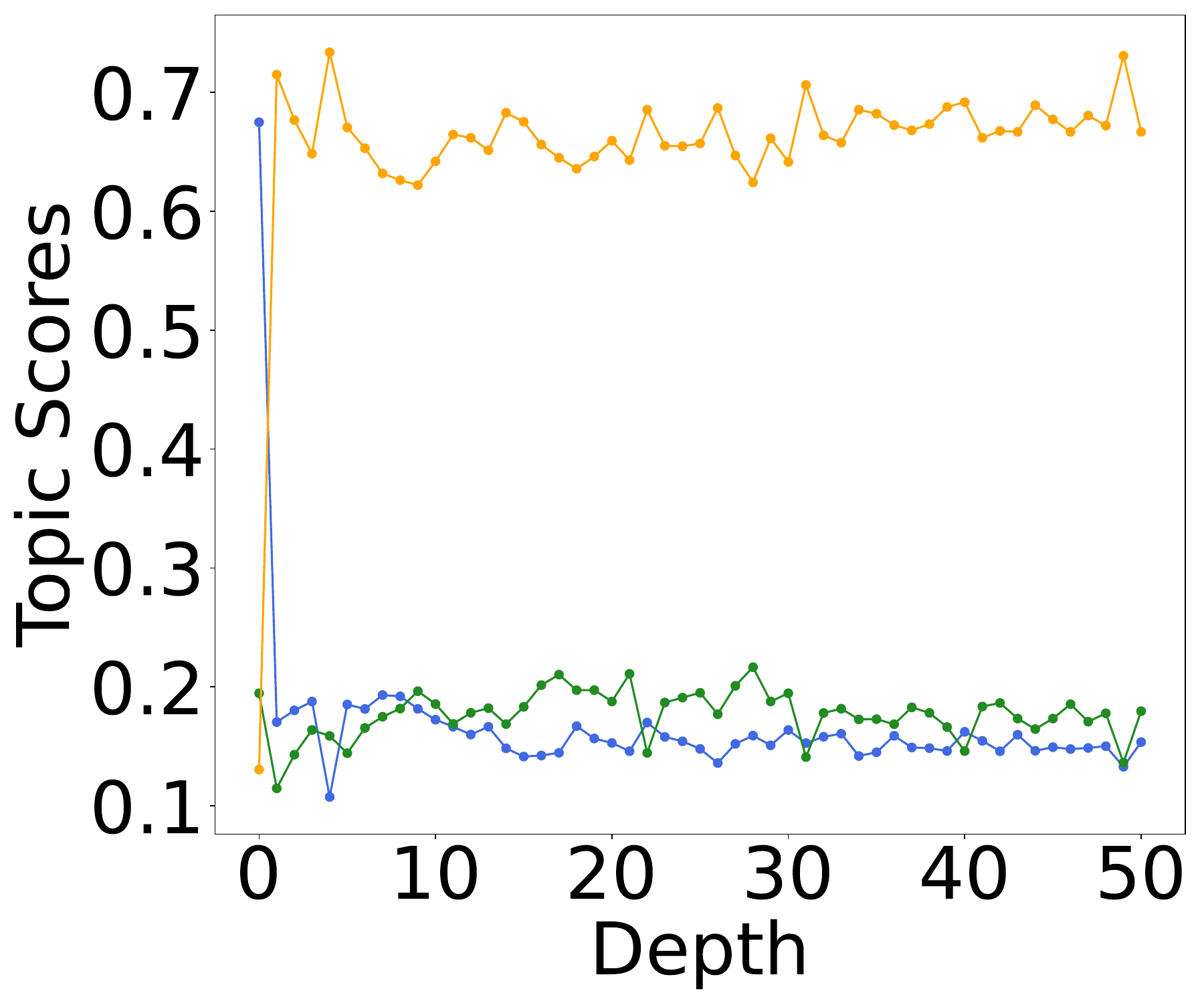}
    \caption{\centering \shortstack{SCS Topic Levels\\(15 Secs)}}
    \label{scs-topic_15}
  \end{subfigure}
  \hfill
  \begin{subfigure}[b]{0.32\textwidth}
    \includegraphics[width=\linewidth]{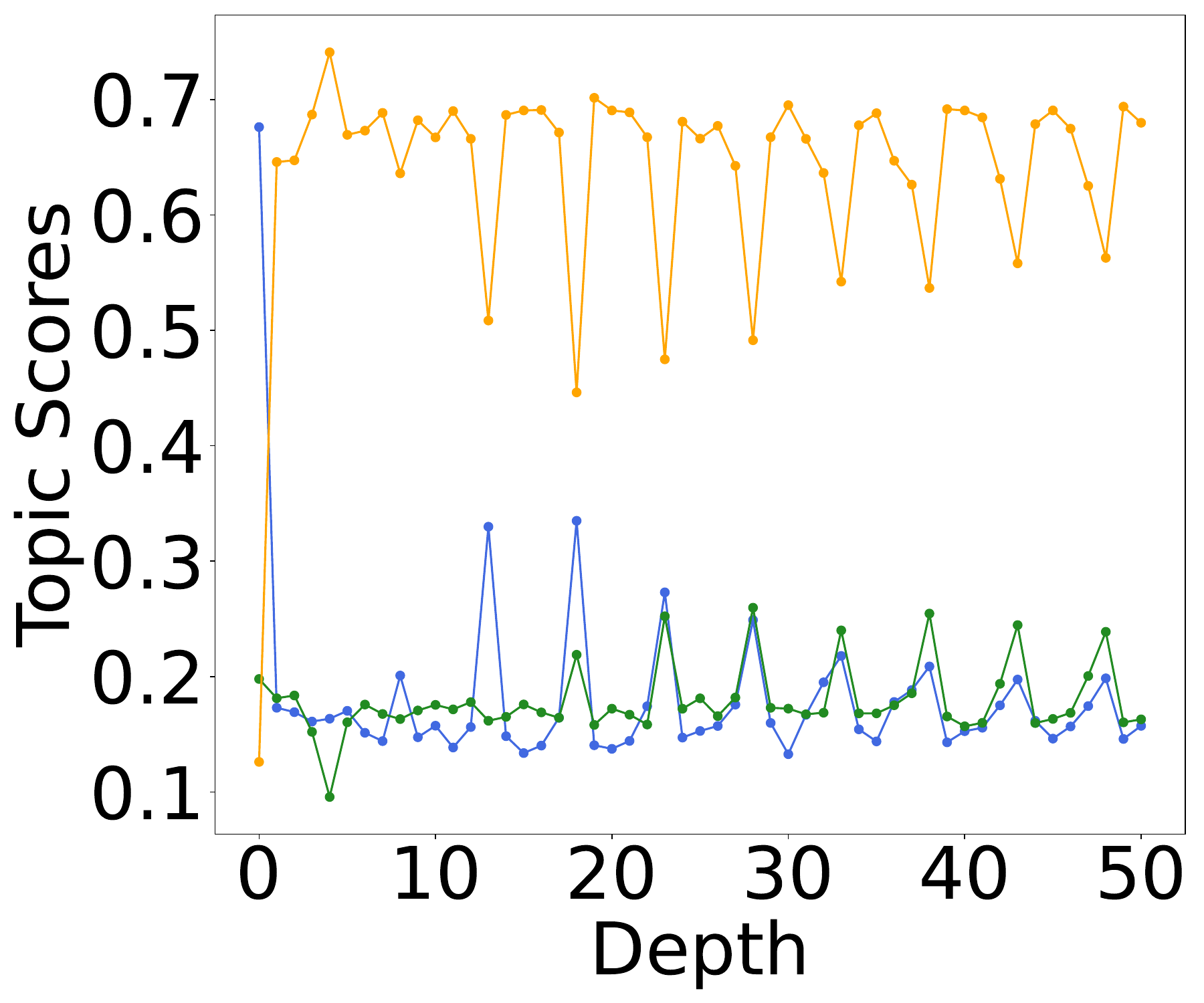}
    \caption{\centering \shortstack{SCS Topic Levels\\(60 Secs)}}
    \label{scs-topic_60}
  \end{subfigure}

  \vspace{0cm}

  \begin{subfigure}[b]{0.32\textwidth}
    \includegraphics[width=\linewidth]{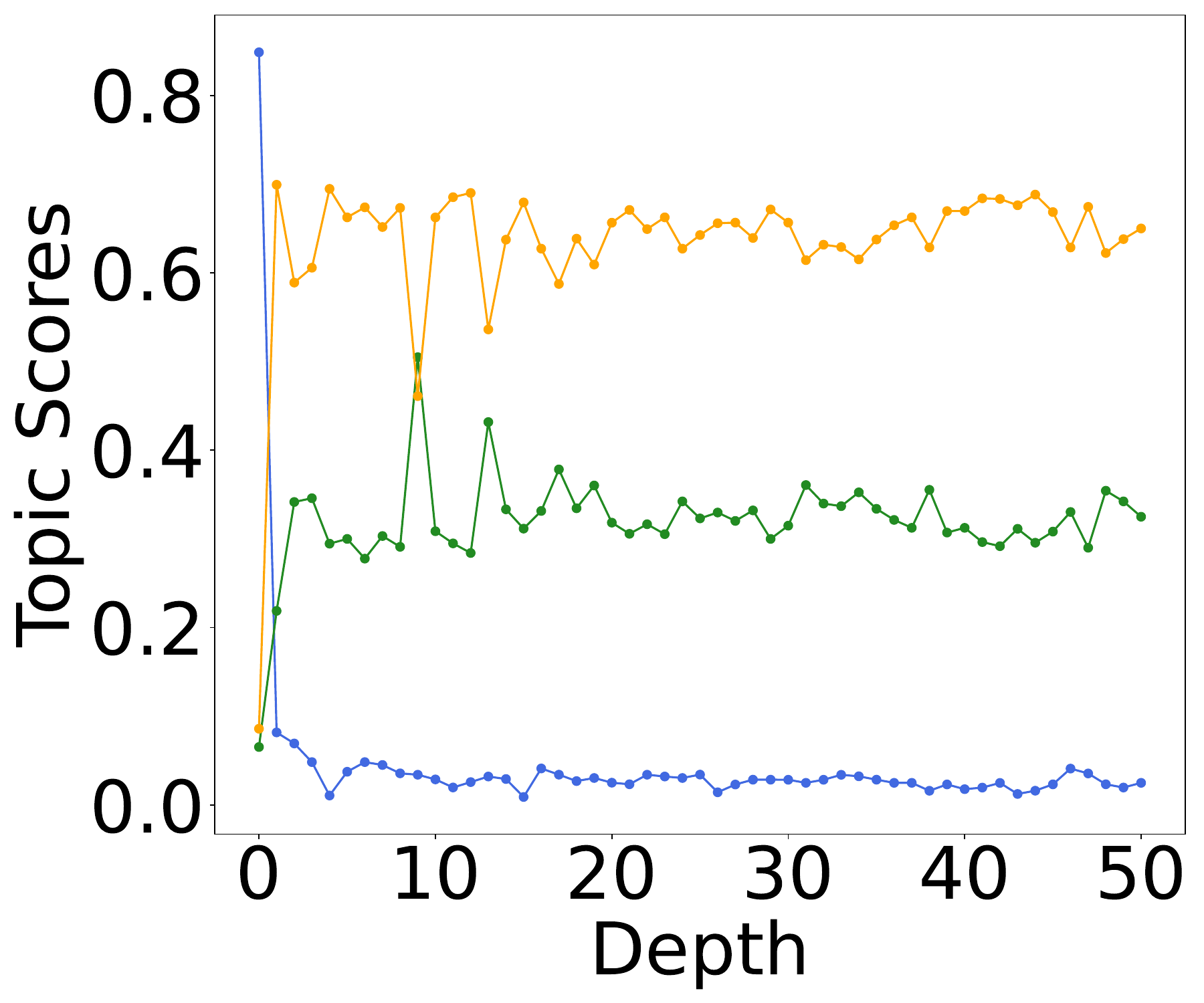}
    \caption{\centering \shortstack{Taiwan Topic Levels\\(3 Secs)}}
    \label{taiwan-topic_3}
  \end{subfigure}
  \hfill
  \begin{subfigure}[b]{0.32\textwidth}
    \includegraphics[width=\linewidth]{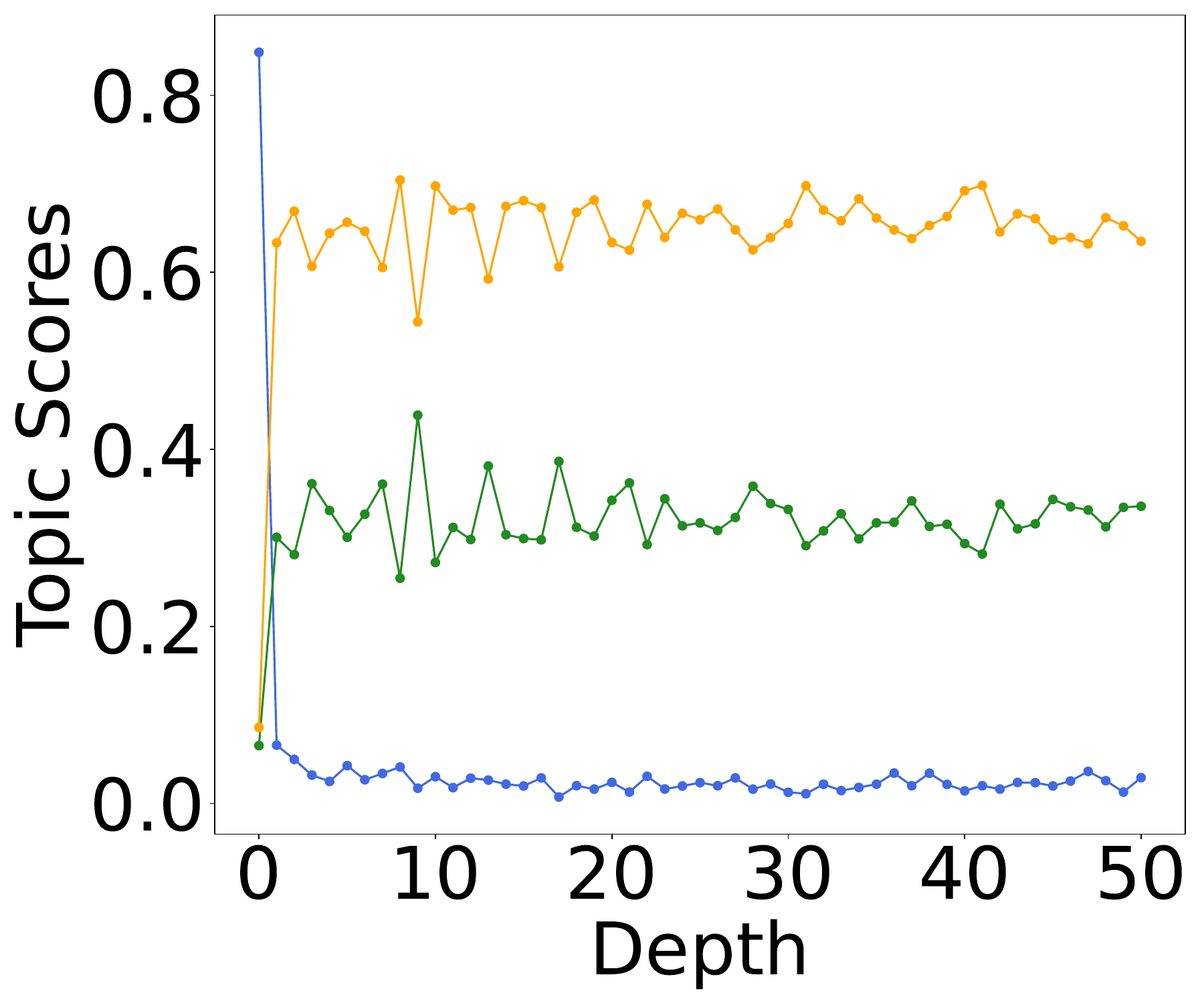}
    \caption{\centering \shortstack{Taiwan Topic Levels\\(15 Secs)}}
    \label{taiwan-topic_15}
  \end{subfigure}
  \hfill
  \begin{subfigure}[b]{0.32\textwidth}
    \includegraphics[width=\linewidth]{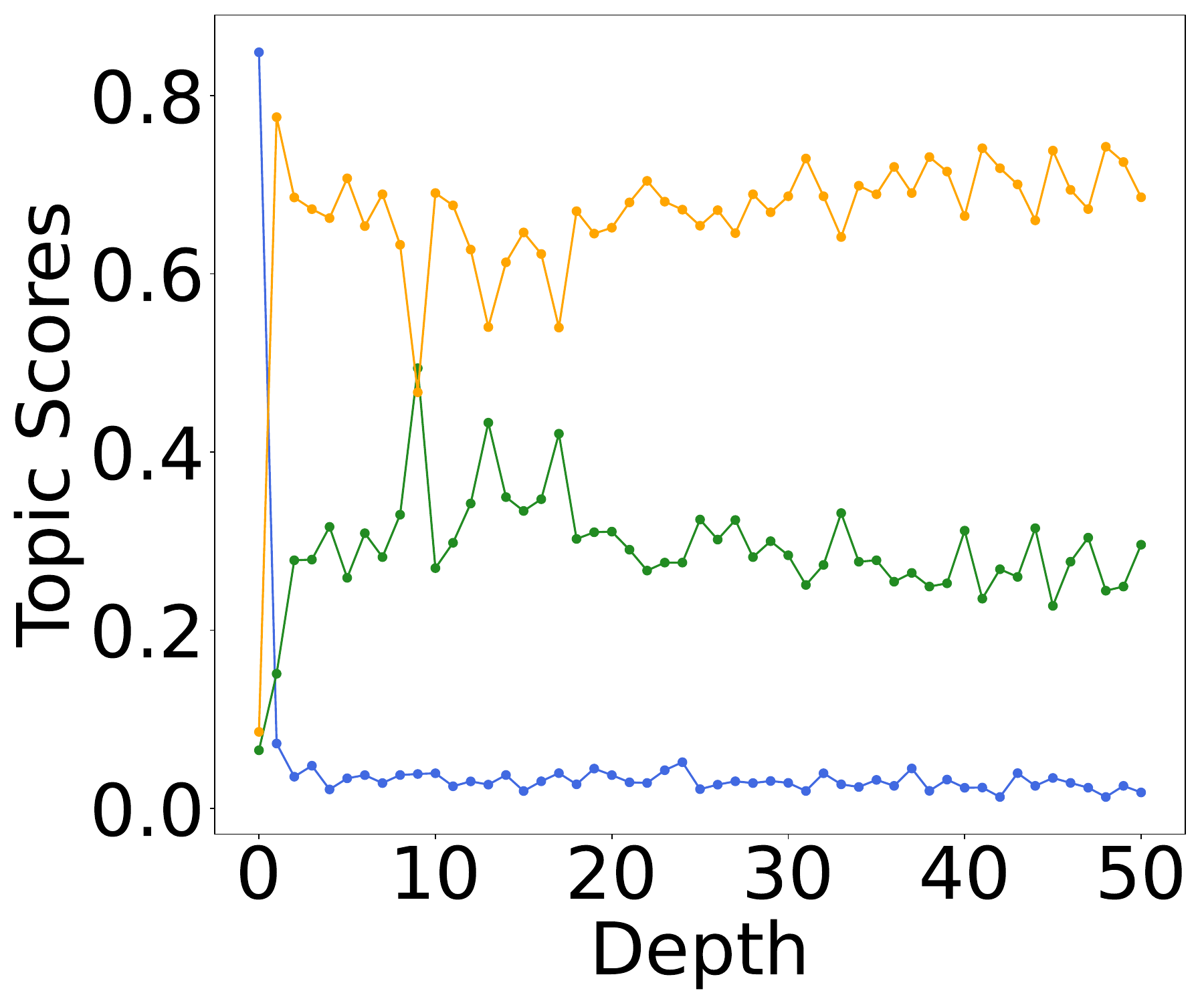}
    \caption{\centering \shortstack{Taiwan Topic Levels\\(60 Secs)}}
    \label{taiwan-topic_60}
  \end{subfigure}

  \vspace{0cm}

  \begin{subfigure}[b]{0.32\textwidth}
    \includegraphics[width=\linewidth]{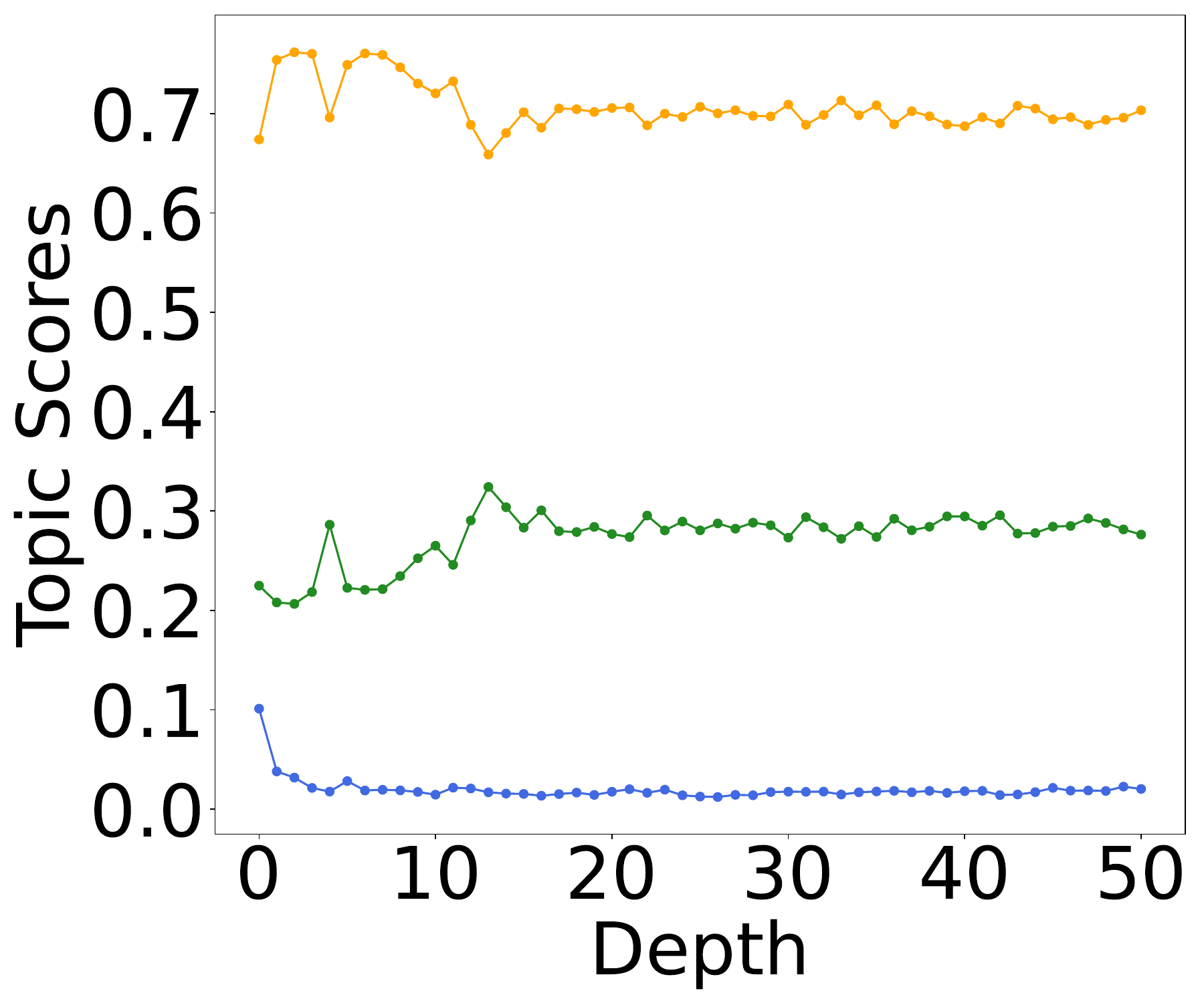}
    \caption{\centering \shortstack{General Topic Levels\\(3 Secs)}}
    \label{yt-general-topic_3}
  \end{subfigure}
  \hfill
  \begin{subfigure}[b]{0.32\textwidth}
    \includegraphics[width=\linewidth]{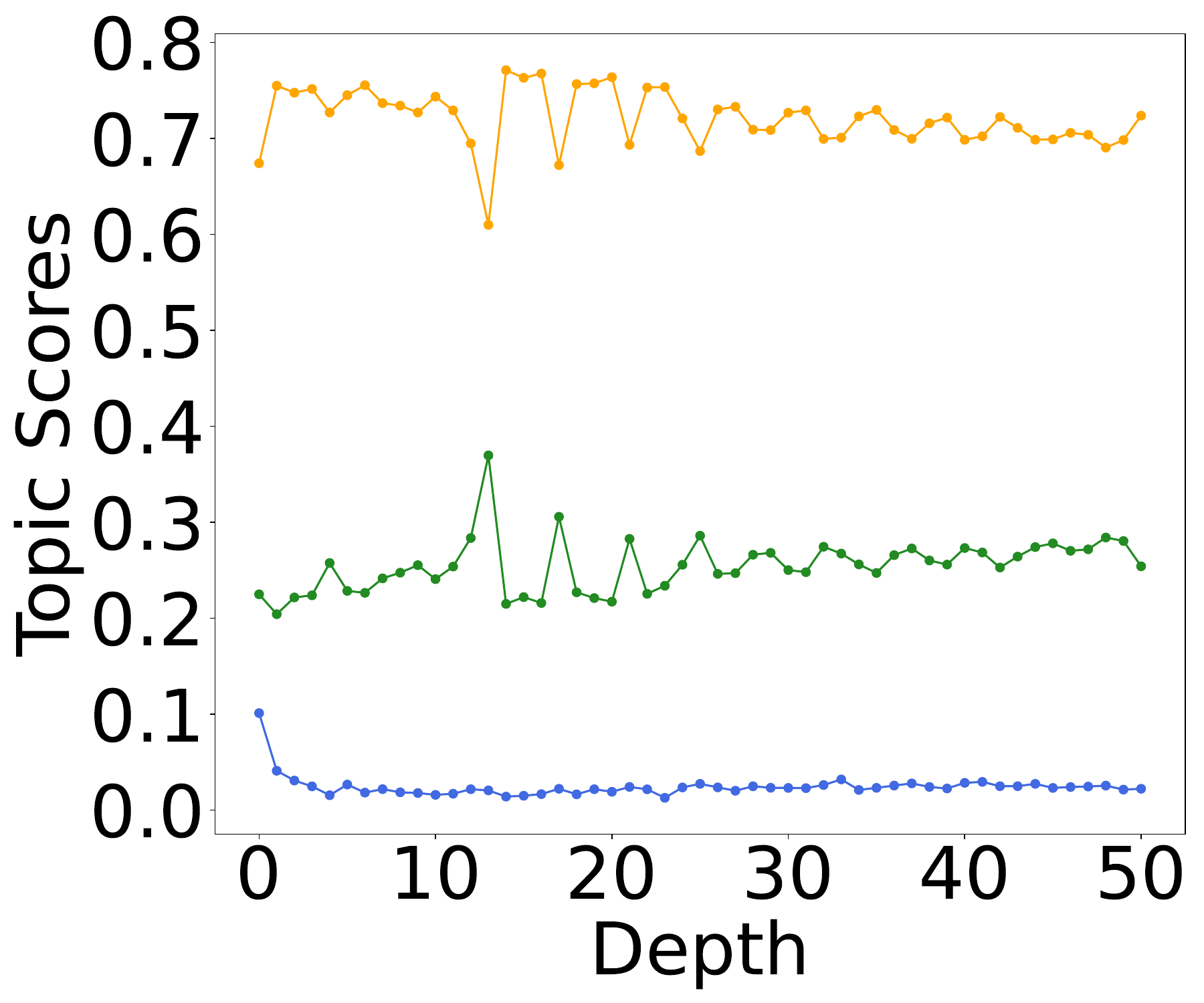}
    \caption{\centering \shortstack{General Topic Levels\\(15 Secs)}}
    \label{yt-general-topic_15}
  \end{subfigure}
  \hfill
  \begin{subfigure}[b]{0.32\textwidth}
    \includegraphics[width=\linewidth]{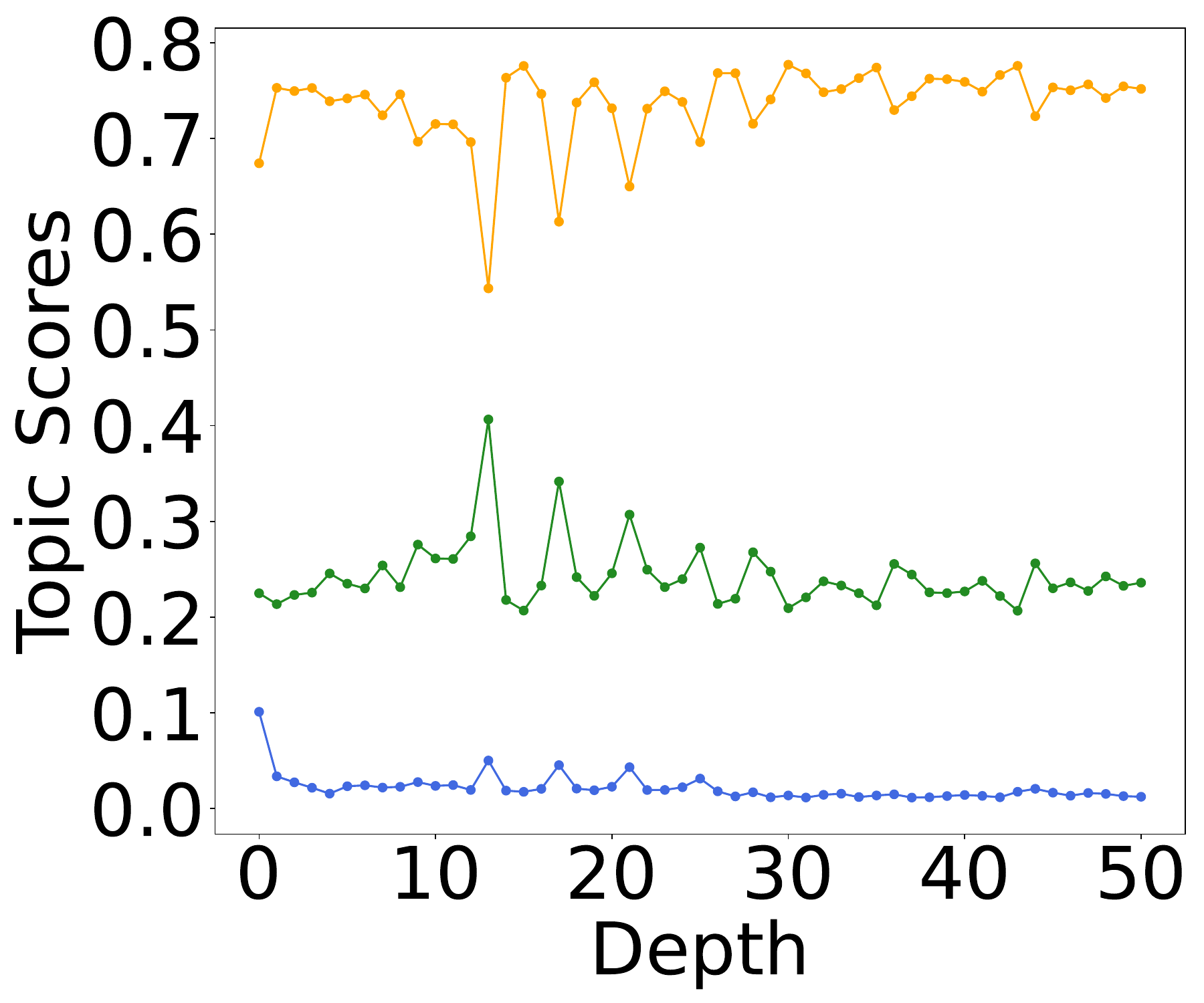}
    \caption{\centering \shortstack{General Topic Levels\\(60 Secs)}}
    \label{yt-general-topic_60}
  \end{subfigure}

  \caption{Topic distribution across South China Sea (SCS), Taiwan Election, and General YouTube Content videos at each recommendation depth. The different lines represent topics such as entertainment, non-entertainment, and politics, and show the mean values at each depth.}

  \label{fig:all_topic}
\end{figure}

To further analyze content drift, we examined the distribution of recommended video topics across three categories: \textit{politics}, \textit{non-entertainment}, and \textit{entertainment}. Topic classification was applied to all datasets using the validated generative AI model, with results shown in Figure~\ref{fig:all_topic}. In the South China Sea (Figures~\ref{scs-topic_3}, \ref{scs-topic_15}, \ref{scs-topic_60}) and Taiwan Election datasets (Figures~\ref{taiwan-topic_3}, \ref{taiwan-topic_15}, \ref{taiwan-topic_60}), initial seed videos (depth 0) were predominantly political, as expected. However, political content declined sharply as recommendation depth increased. In the South China Sea case, entertainment content dominated from depth 1 onward. For Taiwan, both entertainment and non-entertainment topics emerged, with entertainment comprising the majority beyond the first recommendation. This reflects substantial topic drift, where recommendations quickly shift away from politically sensitive content.

In contrast, the General Content dataset (Figures~\ref{yt-general-topic_3}, \ref{yt-general-topic_15}, \ref{yt-general-topic_60}) showed greater topical stability. Entertainment was already dominant at depth 0 and remained so across depths. Notably, political content dropped from a small baseline (around 0.1) to nearly zero, suggesting systematic avoidance of political topics in general recommendations.

These findings indicate a potential algorithmic bias in how YouTube Shorts handles politically sensitive content, favoring entertainment over thematic consistency. This likely reflects engagement optimization and popularity bias, as entertainment typically attracts more viewer attention. Watch-time effects were most notable in the full-duration (60-second) condition. As shown in Figures~\ref{scs-topic_60}, \ref{taiwan-topic_60}, and \ref{yt-general-topic_60}, we observed greater topic shifts at deeper depths, particularly beyond the tenth recommendation. In several cases, ads or sponsored content appeared, suggesting monetized content may be strategically timed to coincide with sustained user engagement. In contrast, the 3-second and 15-second watch-time conditions yielded more mixed patterns with no consistent topic transitions.

Overall, topic distribution analysis reveals a consistent drift from political to entertainment content, especially in politically sensitive datasets. In full watch-time conditions, deeper depths showed increased variability and more frequent appearance of promotional content. This suggests YouTube may adjust its recommendation strategy in response to higher engagement, potentially to diversify content or optimize monetization without risking early user drop-off.

\subsection{Emotion Distribution}

\begin{figure}[!ht]
  \centering

  \begin{subfigure}[b]{0.32\textwidth}
    \includegraphics[width=\linewidth]{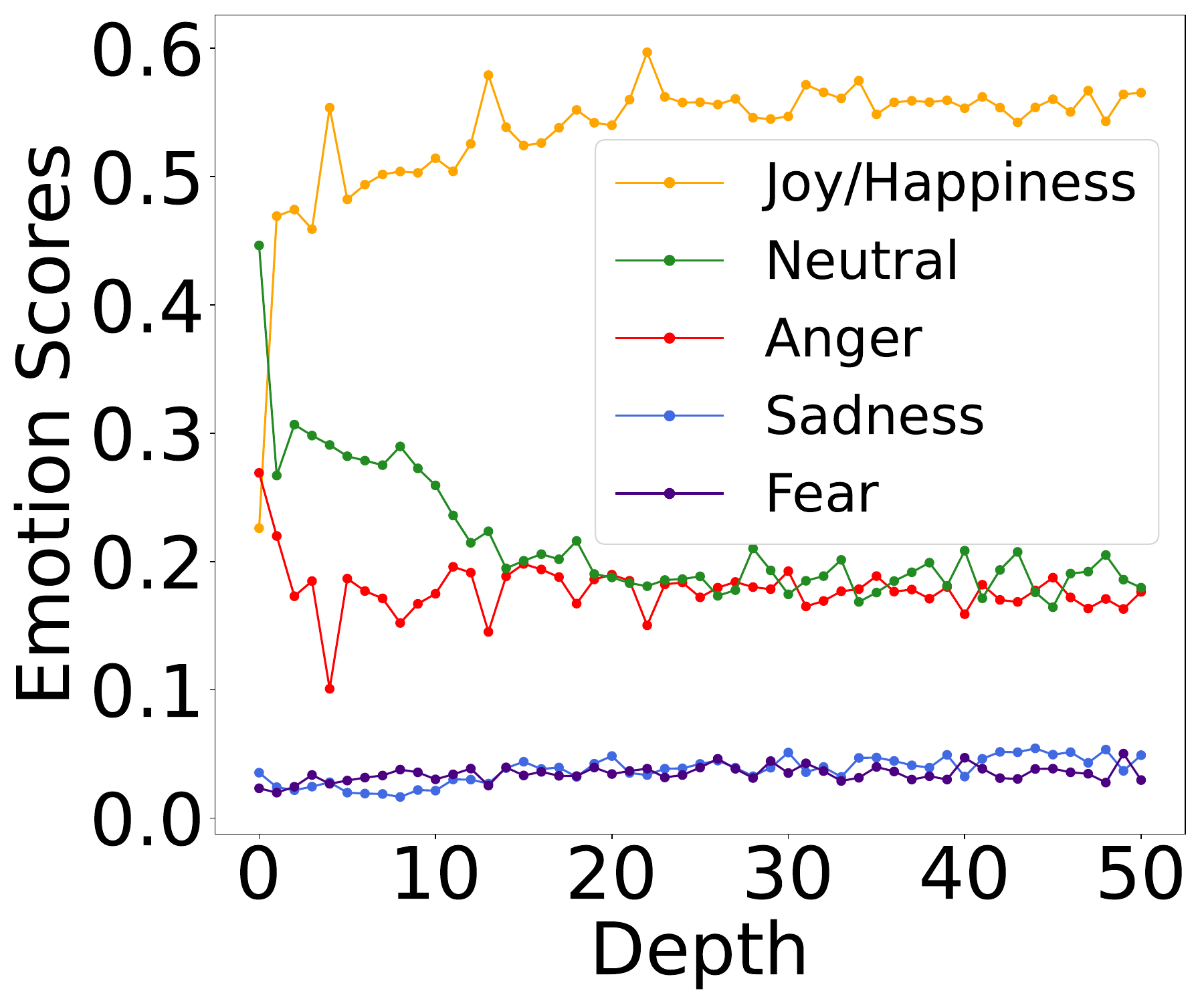}
    \caption{\centering \shortstack{SCS Emotion Levels\\(3 Secs)}}
    \label{scs-emotion_3}
  \end{subfigure}
  \hfill
  \begin{subfigure}[b]{0.32\textwidth}
    \includegraphics[width=\linewidth]{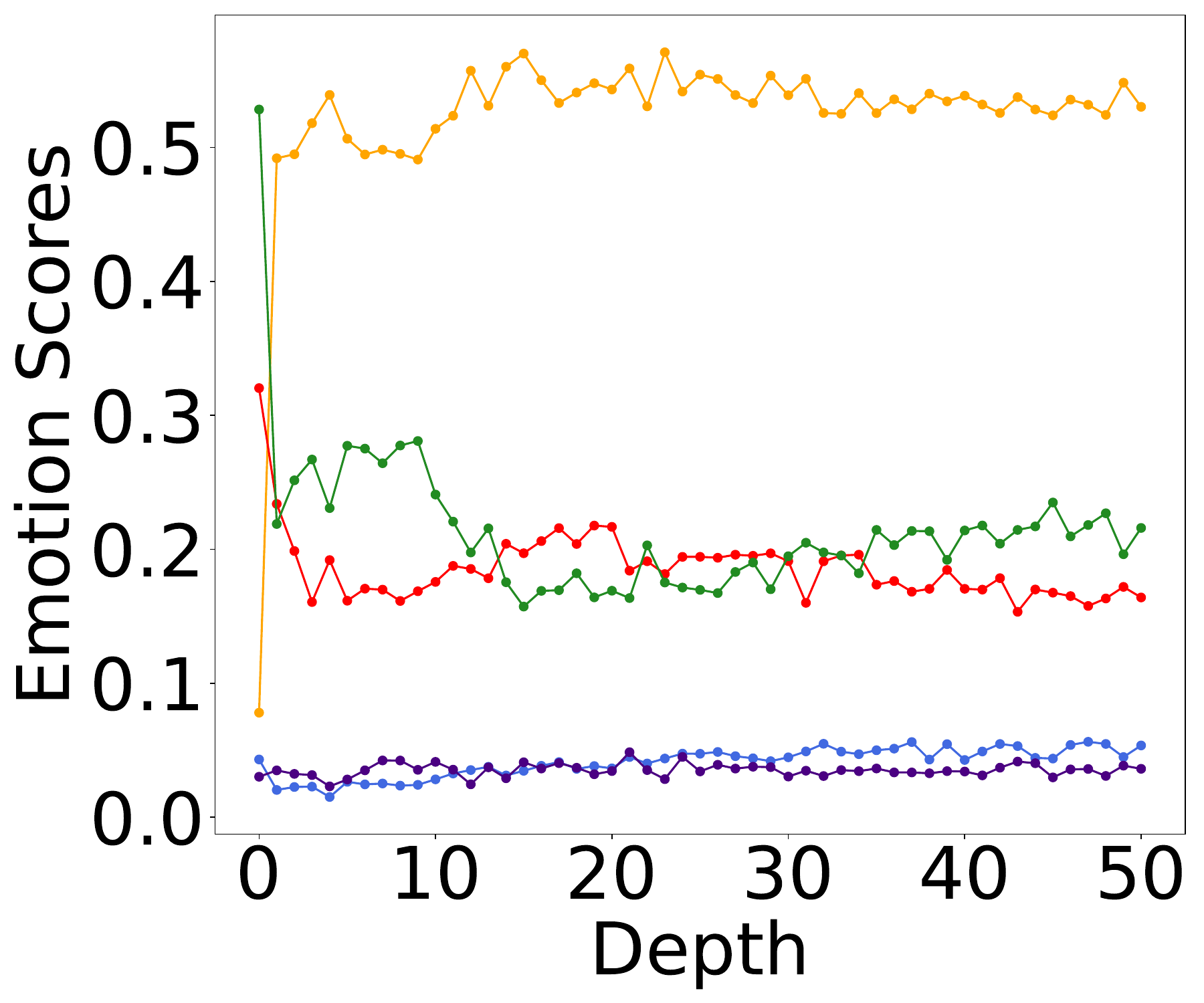}
    \caption{\centering \shortstack{SCS Emotion Levels\\(15 Secs)}}
    \label{scs-emotion_15}
  \end{subfigure}
  \hfill
  \begin{subfigure}[b]{0.32\textwidth}
    \includegraphics[width=\linewidth]{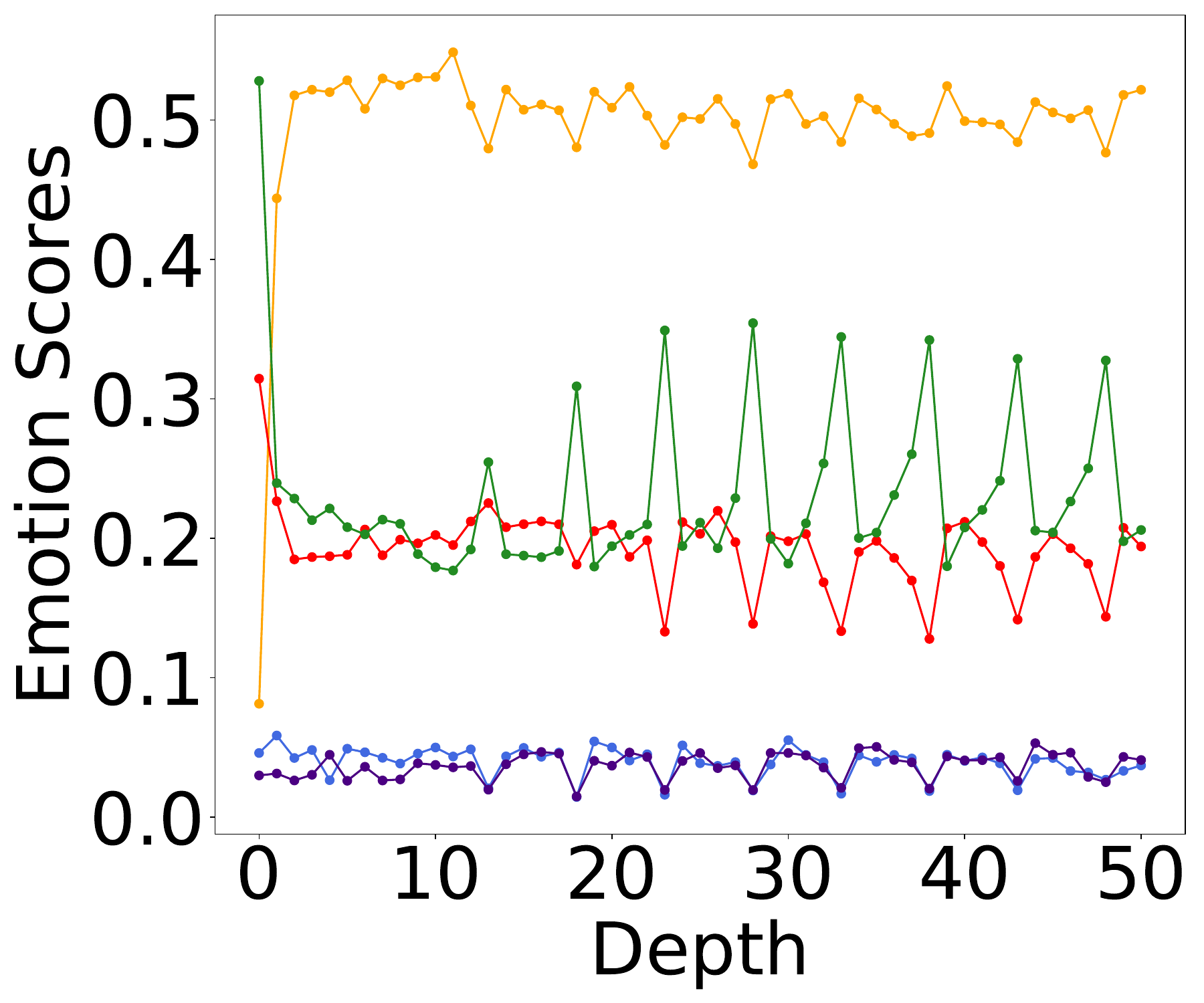}
    \caption{\centering \shortstack{SCS Emotion Levels\\(60 Secs)}}
    \label{scs-emotion_60}
  \end{subfigure}

  \vspace{0cm}

  \begin{subfigure}[b]{0.32\textwidth}
    \includegraphics[width=\linewidth]{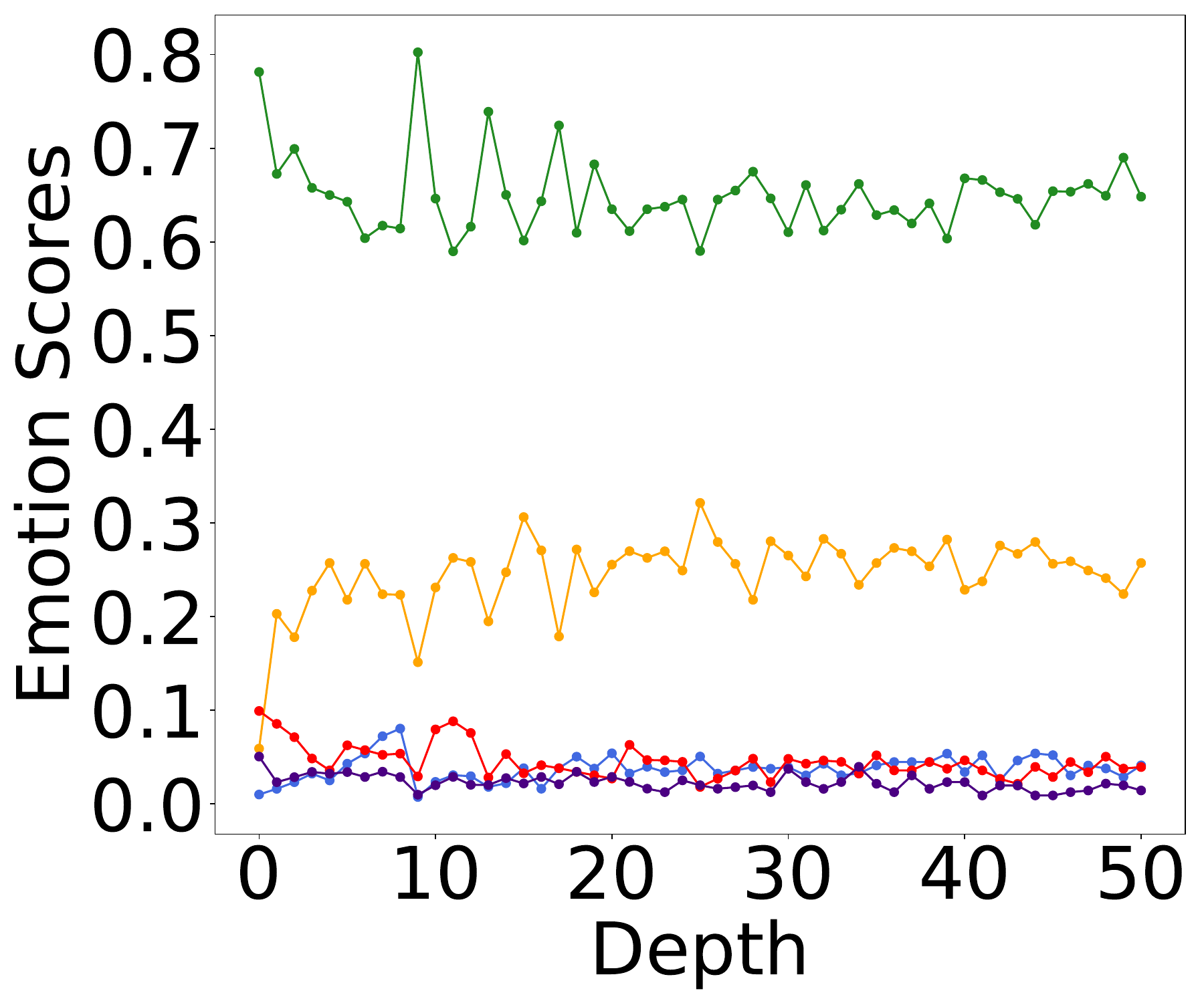}
    \caption{\centering \shortstack{Taiwan Emotion Levels\\(3 Secs)}}
    \label{taiwan-emotion_3}
  \end{subfigure}
  \hfill
  \begin{subfigure}[b]{0.32\textwidth}
    \includegraphics[width=\linewidth]{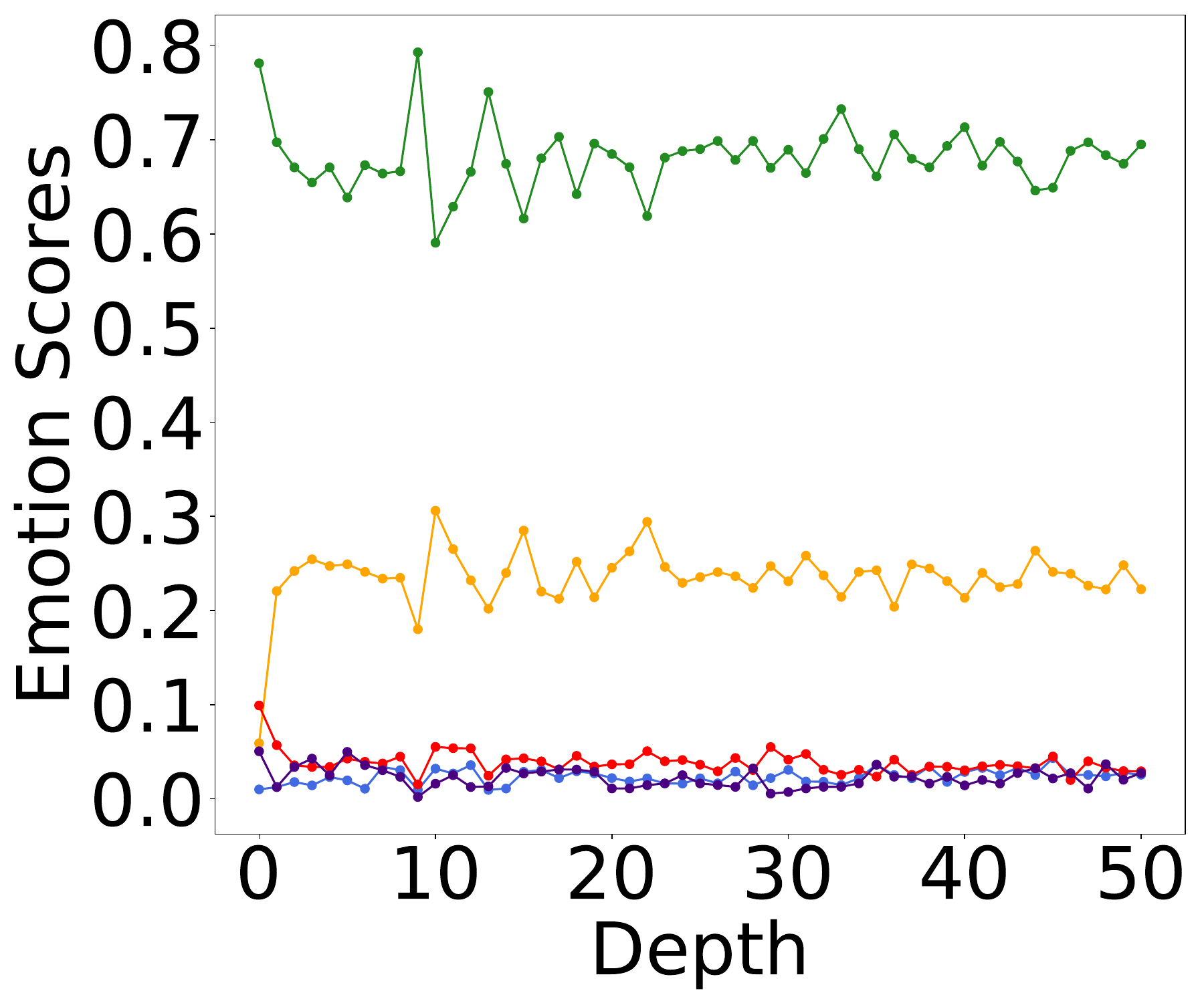}
    \caption{\centering \shortstack{Taiwan Emotion Levels\\(15 Secs)}}
    \label{taiwan-emotion_15}
  \end{subfigure}
  \hfill
  \begin{subfigure}[b]{0.32\textwidth}
    \includegraphics[width=\linewidth]{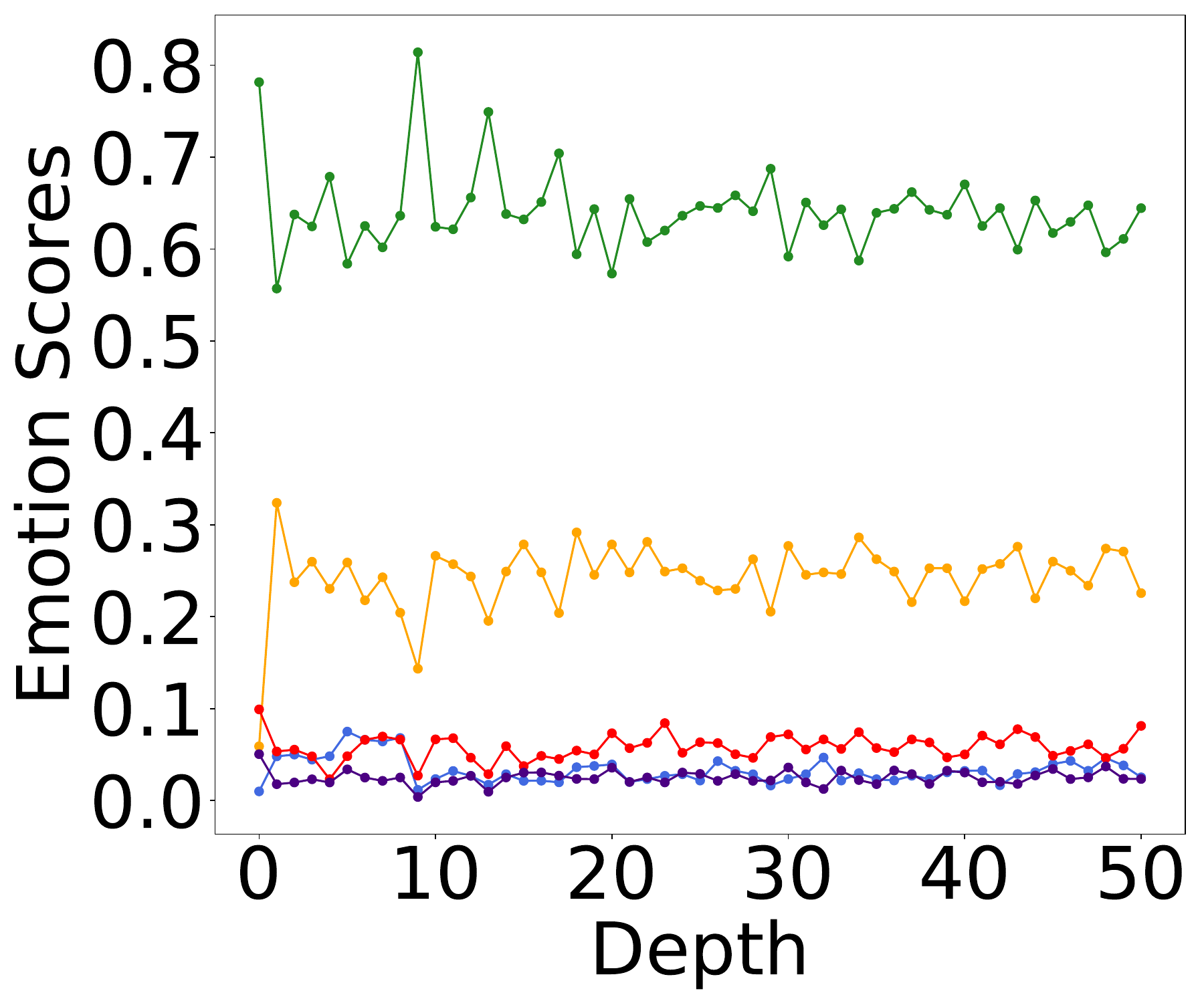}
    \caption{\centering \shortstack{Taiwan Emotion Levels\\(60 Secs)}}
    \label{taiwan-emotion_60}
  \end{subfigure}

\vspace{0cm}
    
  \begin{subfigure}[b]{0.32\textwidth}
    \includegraphics[width=\linewidth]{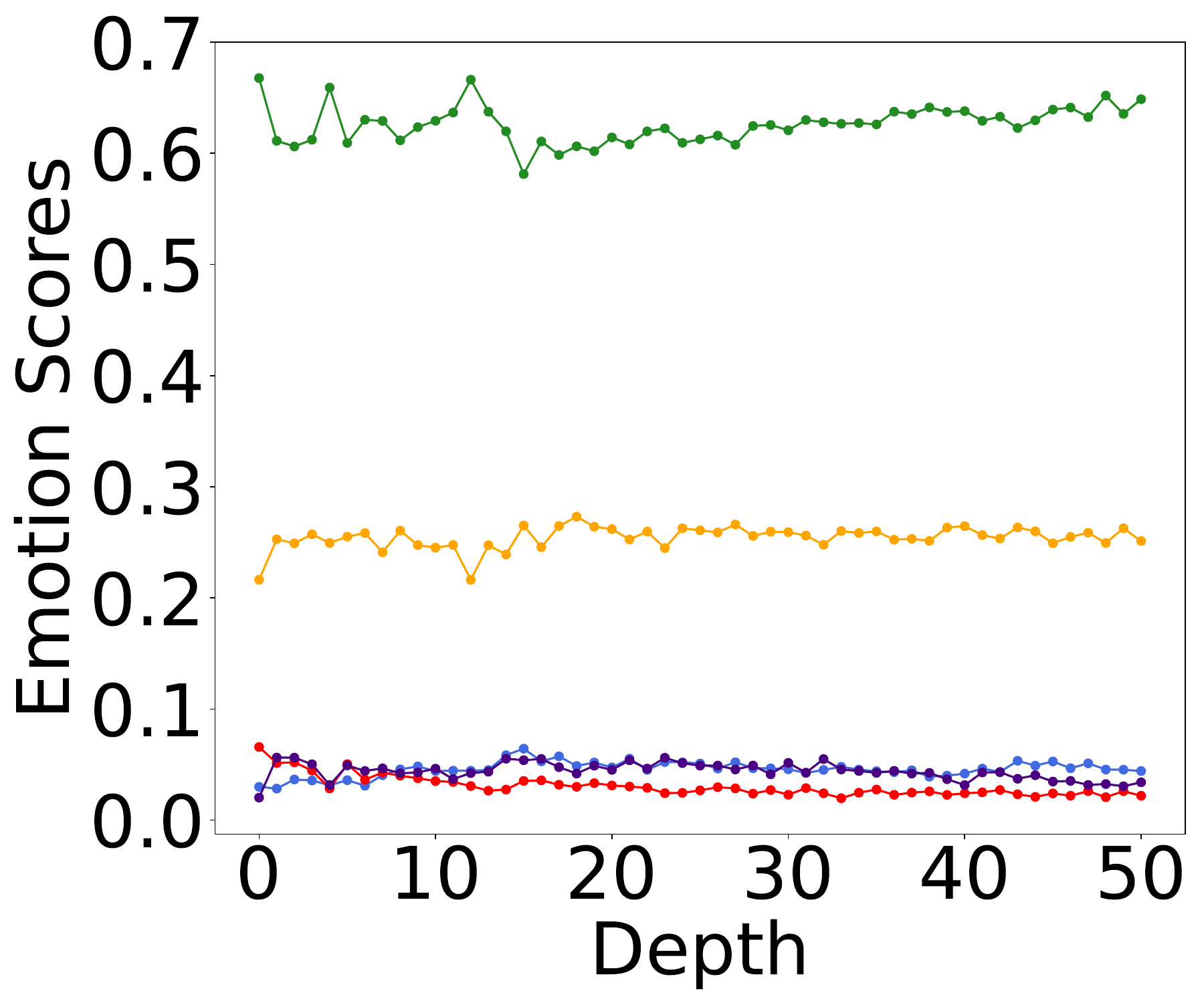}
    \caption{\centering \shortstack{General Emotion Levels\\(3 Secs)}}
    \label{yt-general-emotion_3}
  \end{subfigure}
  \hfill
  \begin{subfigure}[b]{0.32\textwidth}
    \includegraphics[width=\linewidth]{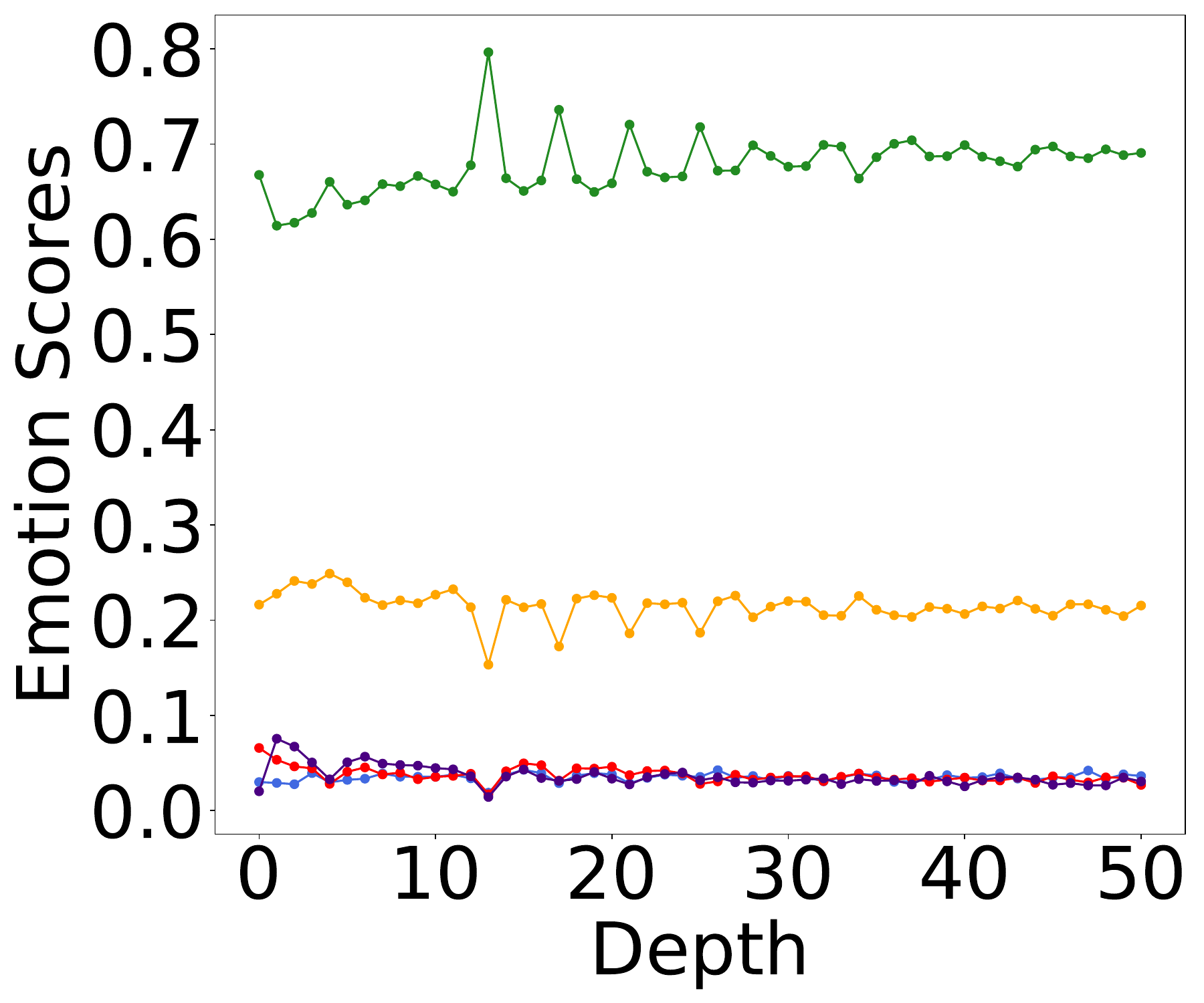}
    \caption{\centering \shortstack{General Emotion Levels\\(15 Secs)}}
    \label{yt-general-emotion_15}
  \end{subfigure}
  \hfill
  \begin{subfigure}[b]{0.32\textwidth}
    \includegraphics[width=\linewidth]{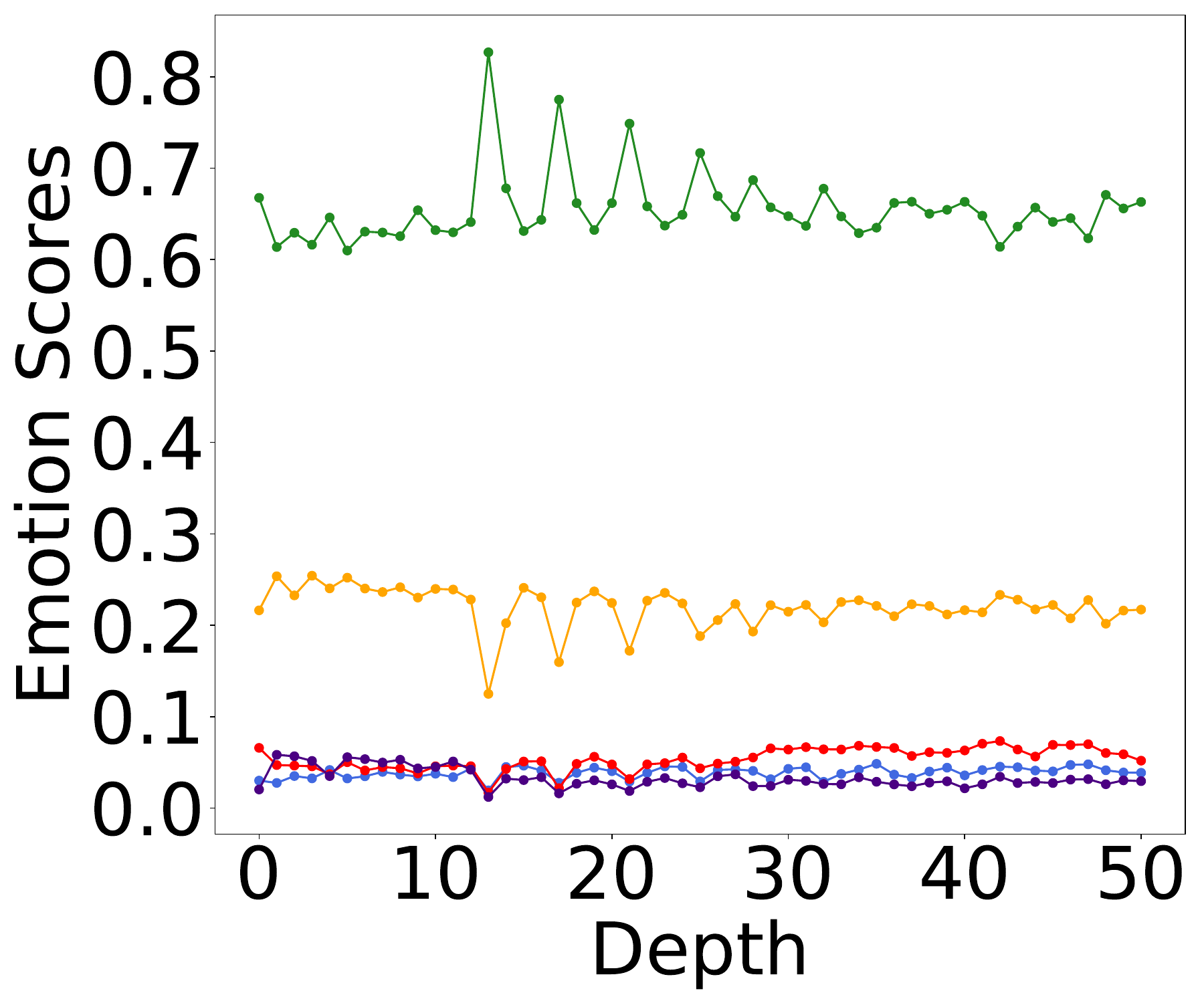}
    \caption{\centering \shortstack{General Emotion Levels\\(60 Secs)}}
    \label{yt-general-emotion_60}
  \end{subfigure}

  \caption{Emotion distribution across South China Sea (SCS), Taiwan Election, and General YouTube Content videos at each recommendation depth. The different lines represent emotions such as joy, neutral, anger, sadness, and fear, and show the mean values at each depth.}

  \label{fig:all_emotion}
\end{figure}

To analyze how emotional tone evolves across YouTube Shorts recommendations, we classified each video into one of five categories: \textit{joy/happiness}, \textit{sadness}, \textit{anger}, \textit{neutral}, and \textit{fear}, as shown in Figure~\ref{fig:all_emotion}. In the South China Sea dataset (Figures~\ref{scs-emotion_3}, \ref{scs-emotion_15}, \ref{scs-emotion_60}), initial seed videos were dominated by \textit{neutral} and \textit{anger}, likely due to the topic’s geopolitical nature. By depth 1, there was a marked increase in \textit{joy/happiness}, with corresponding declines in \textit{neutral} and \textit{anger}, a trend that continued in deeper levels. This suggests the algorithm may steer users toward more emotionally positive content over time.

In the Taiwan Election dataset (Figures~\ref{taiwan-emotion_3}, \ref{taiwan-emotion_15}, \ref{taiwan-emotion_60}), the emotional profile was more subdued. Initial videos were largely \textit{neutral}, with minimal \textit{anger}, likely reflecting the focus on electoral discourse rather than direct conflict. However, similar to the South China Sea case, \textit{joy/happiness} increased after early depths, while negative emotions declined, indicating a consistent preference for emotionally uplifting content even in political narratives.

In contrast, the General YouTube Content dataset showed stable emotional distributions across depths (Figures~\ref{yt-general-emotion_3}, \ref{yt-general-emotion_15}, \ref{yt-general-emotion_60}), with high levels of \textit{joy/happiness} and \textit{neutral} sentiment already present at the seed level, leaving less room for drift.

In the full-duration (60-second) condition, the South China Sea dataset displayed an oscillating emotional pattern (Figure~\ref{scs-emotion_60}), similar to its topic drift (Figure~\ref{scs-topic_60}). These shifts may result from inserted sponsored content, often carrying neutral tones, suggesting YouTube may time such videos during longer viewing sessions to manage affective load and maintain advertiser alignment. Minor variations also appeared in the General Content dataset at 15 and 60 seconds (Figures~\ref{yt-general-emotion_15} and \ref{yt-general-emotion_60}), though without consistent patterns.

Overall, the emotion analysis reveals a clear drift toward positive sentiment in politically sensitive narratives. These trends align with earlier results on relevance and topic, supporting the hypothesis that YouTube Shorts' algorithm shapes both topical direction and emotional tone, potentially to optimize engagement or preserve a consistent platform experience.

\subsection{Engagement Levels}

\begin{figure}[!ht]
  \centering

  \begin{subfigure}[b]{0.32\textwidth}
    \includegraphics[width=\linewidth]{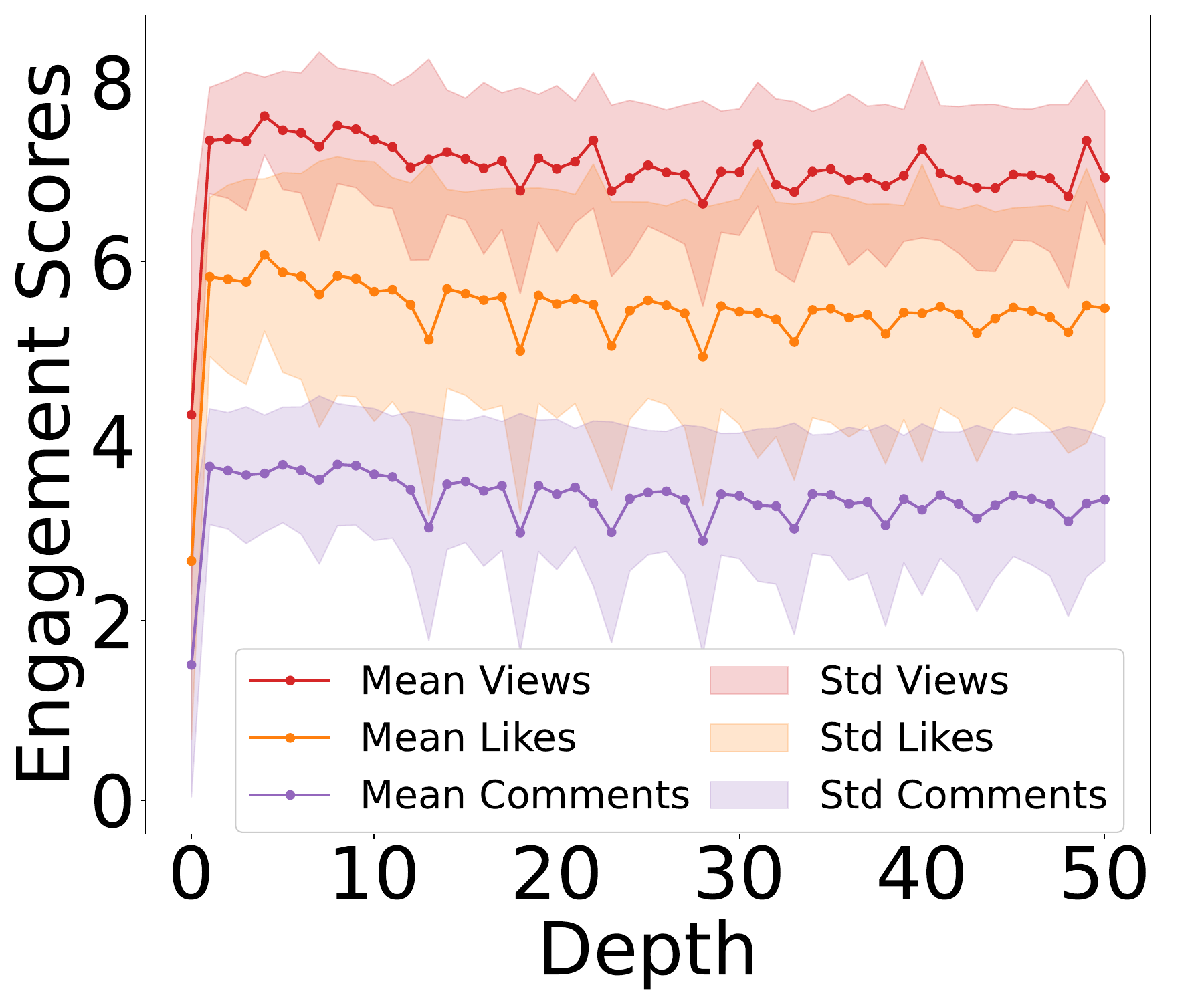}
    \caption{\centering \shortstack{SCS Engagements\\(3 Secs)}}
    \label{scs-engagement_3}
  \end{subfigure}
  \hfill
  \begin{subfigure}[b]{0.32\textwidth}
    \includegraphics[width=\linewidth]{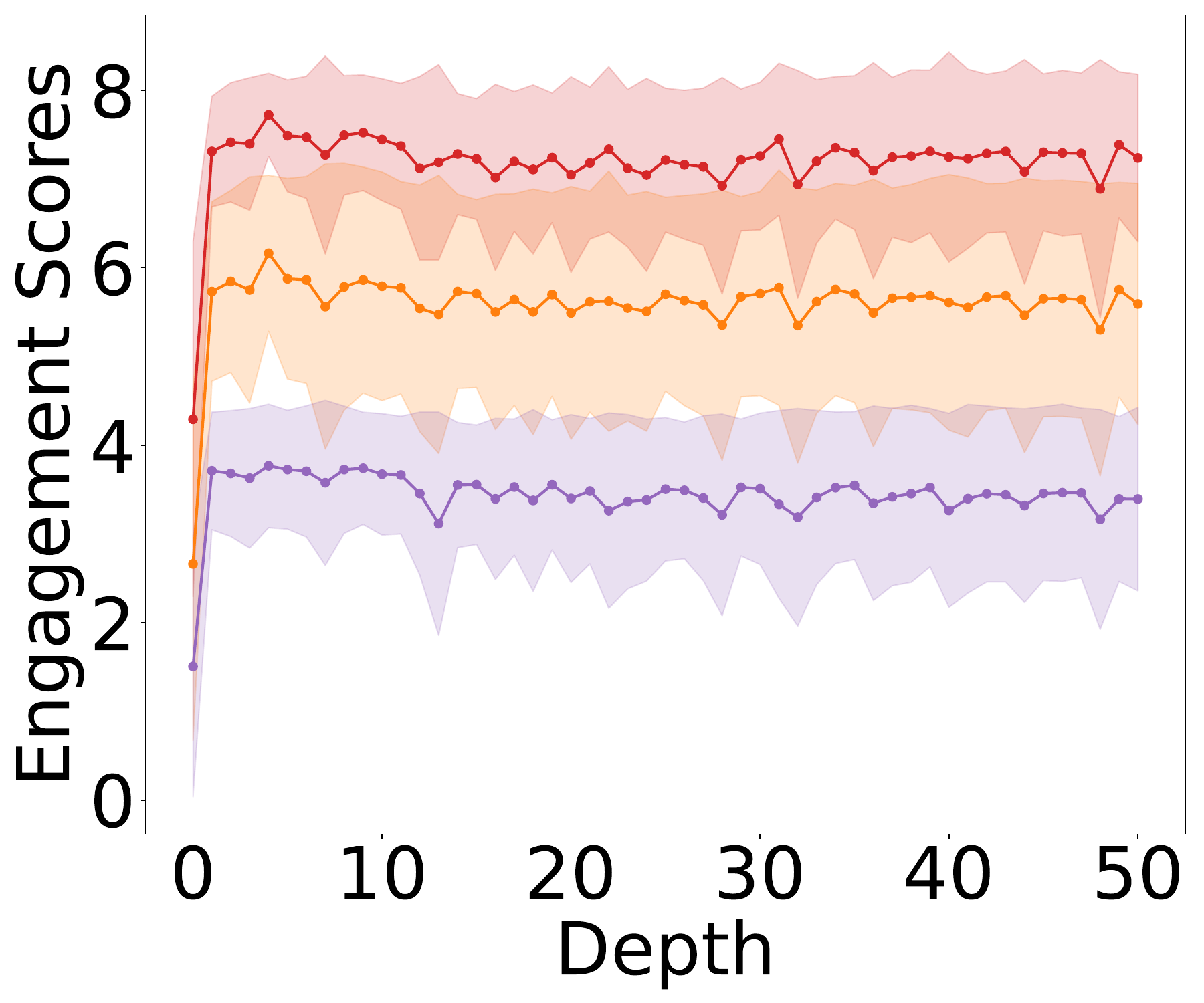}
    \caption{\centering \shortstack{SCS Engagements\\(15 Secs)}}
    \label{scs-engagement_15}
  \end{subfigure}
  \hfill
  \begin{subfigure}[b]{0.32\textwidth}
    \includegraphics[width=\linewidth]{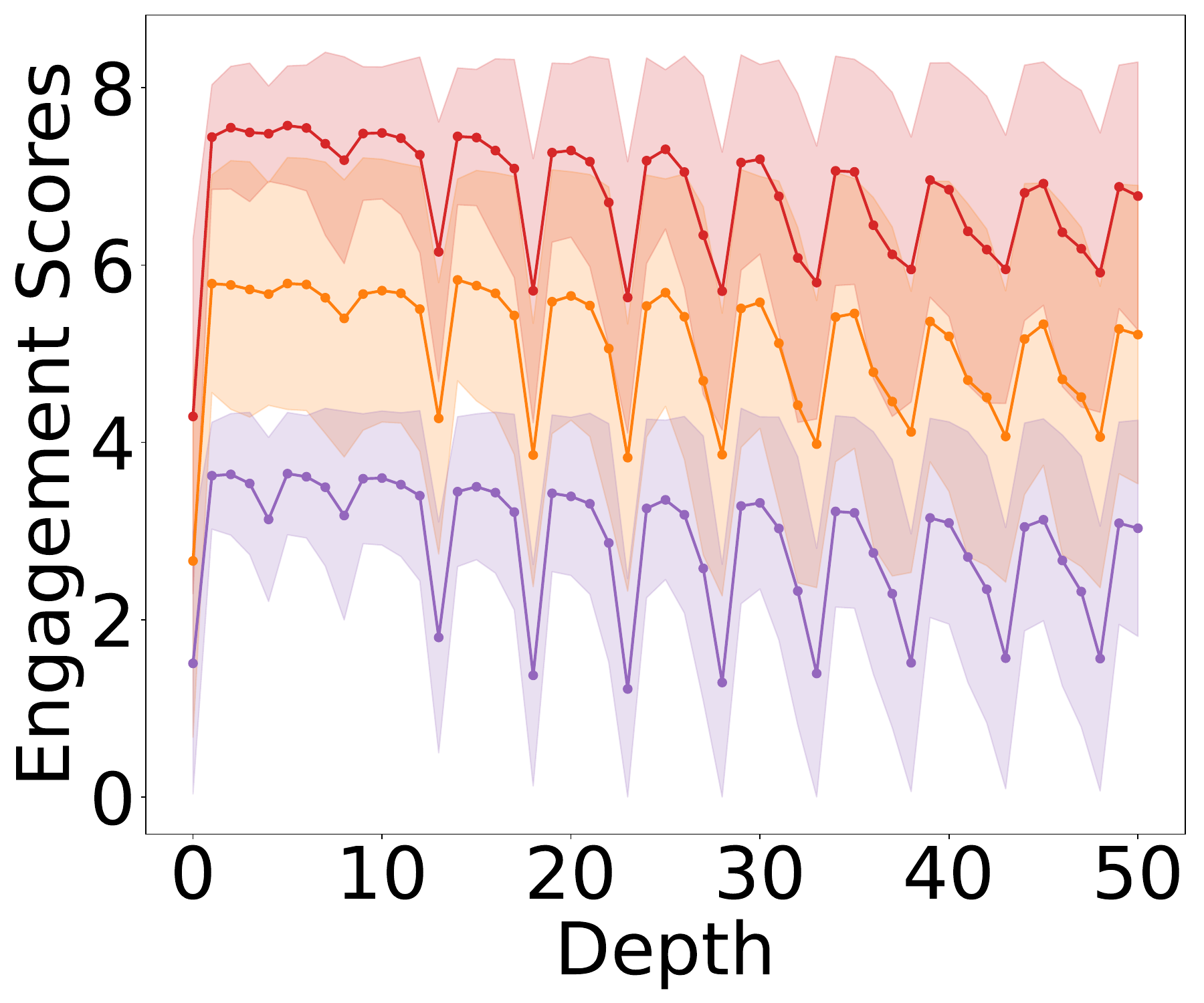}
    \caption{\centering \shortstack{SCS Engagements\\(60 Secs)}}
    \label{scs-engagement_60}
  \end{subfigure}

  \vspace{0cm}

  \begin{subfigure}[b]{0.32\textwidth}
    \includegraphics[width=\linewidth]{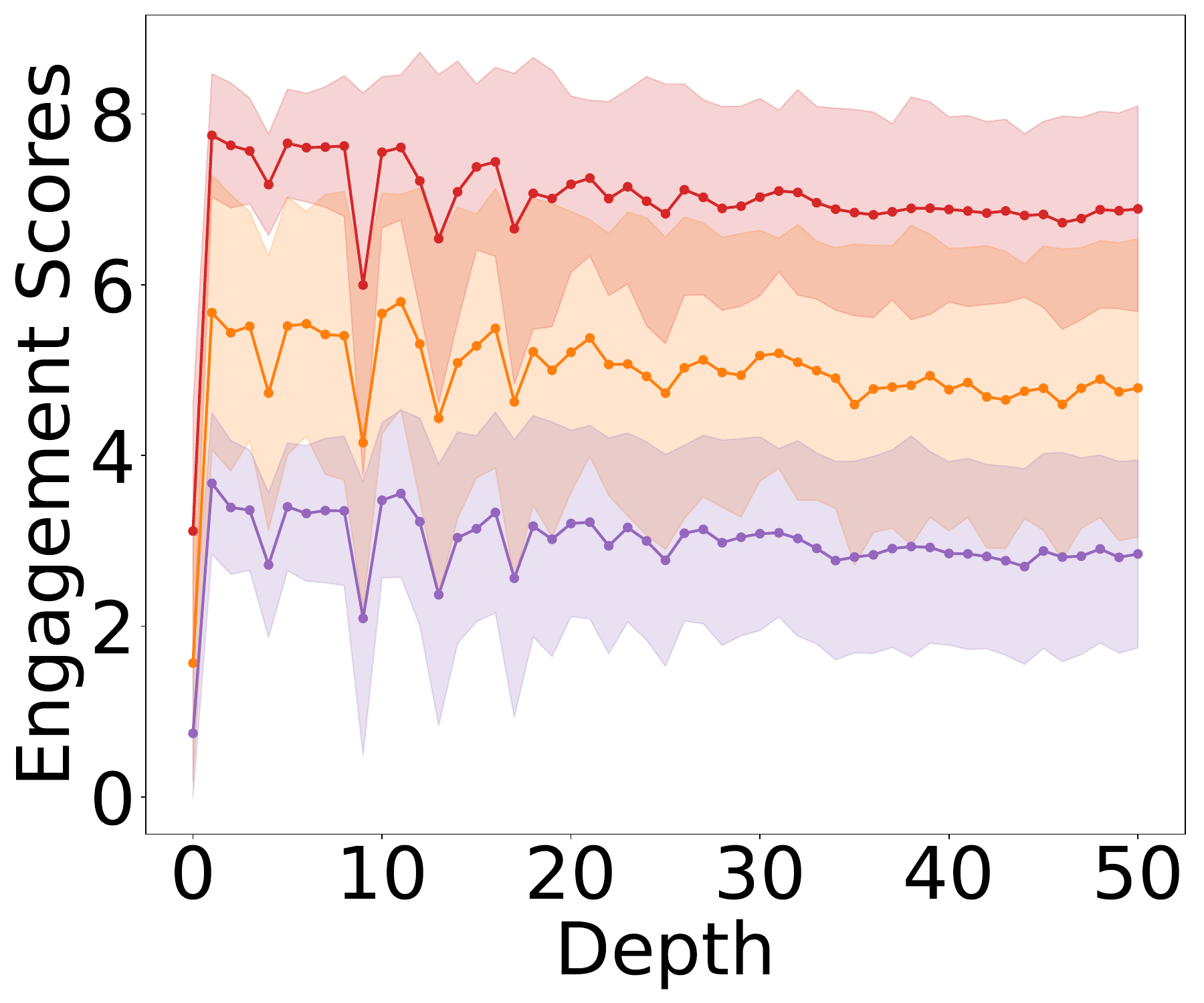}
    \caption{\centering \shortstack{Taiwan Engagements\\(3 Secs)}}
    \label{taiwan-engagement_3}
  \end{subfigure}
  \hfill
  \begin{subfigure}[b]{0.32\textwidth}
    \includegraphics[width=\linewidth]{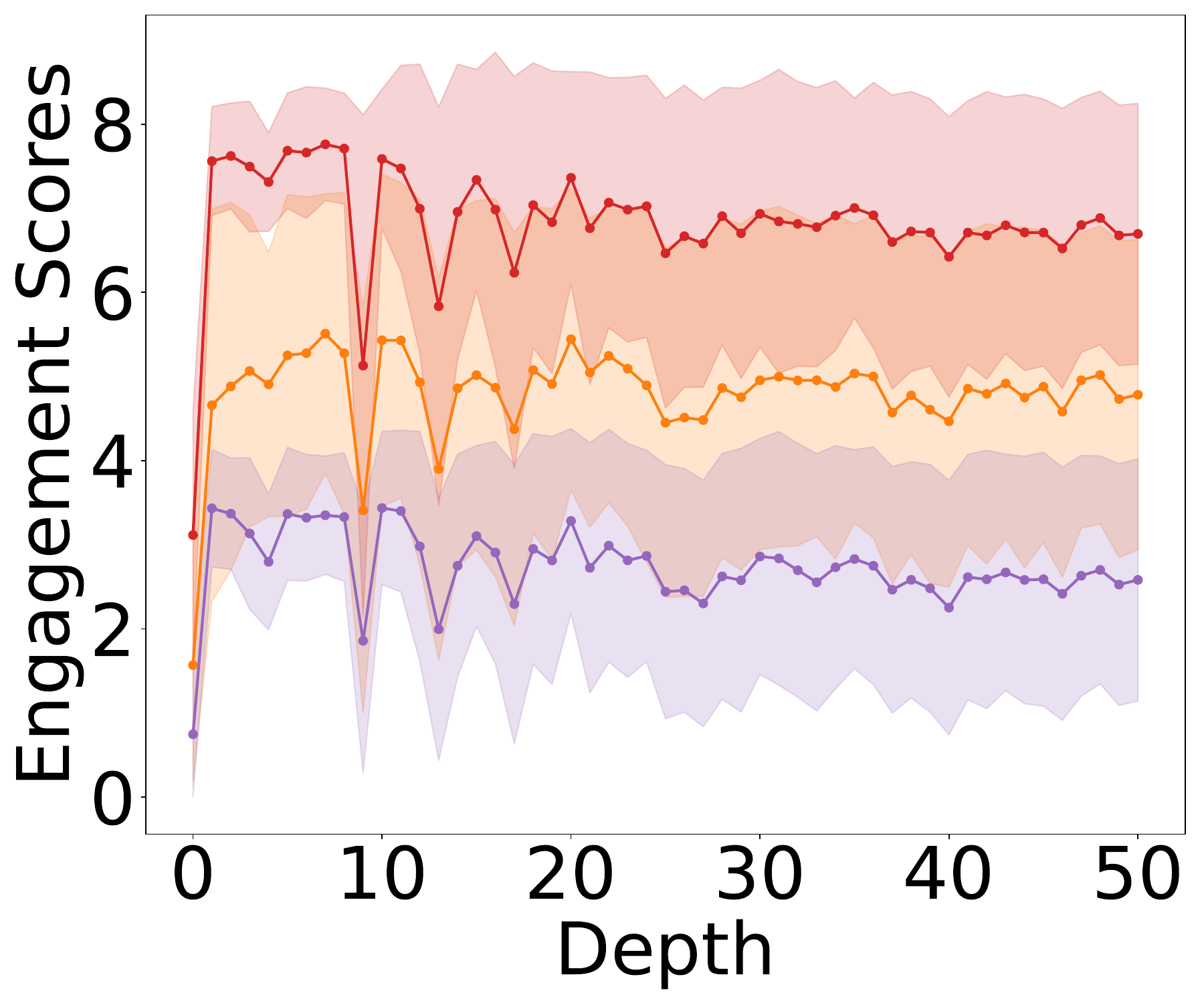}
    \caption{\centering \shortstack{Taiwan Engagements\\(15 Secs)}}
    \label{taiwan-engagement_15}
  \end{subfigure}
  \hfill
  \begin{subfigure}[b]{0.32\textwidth}
    \includegraphics[width=\linewidth]{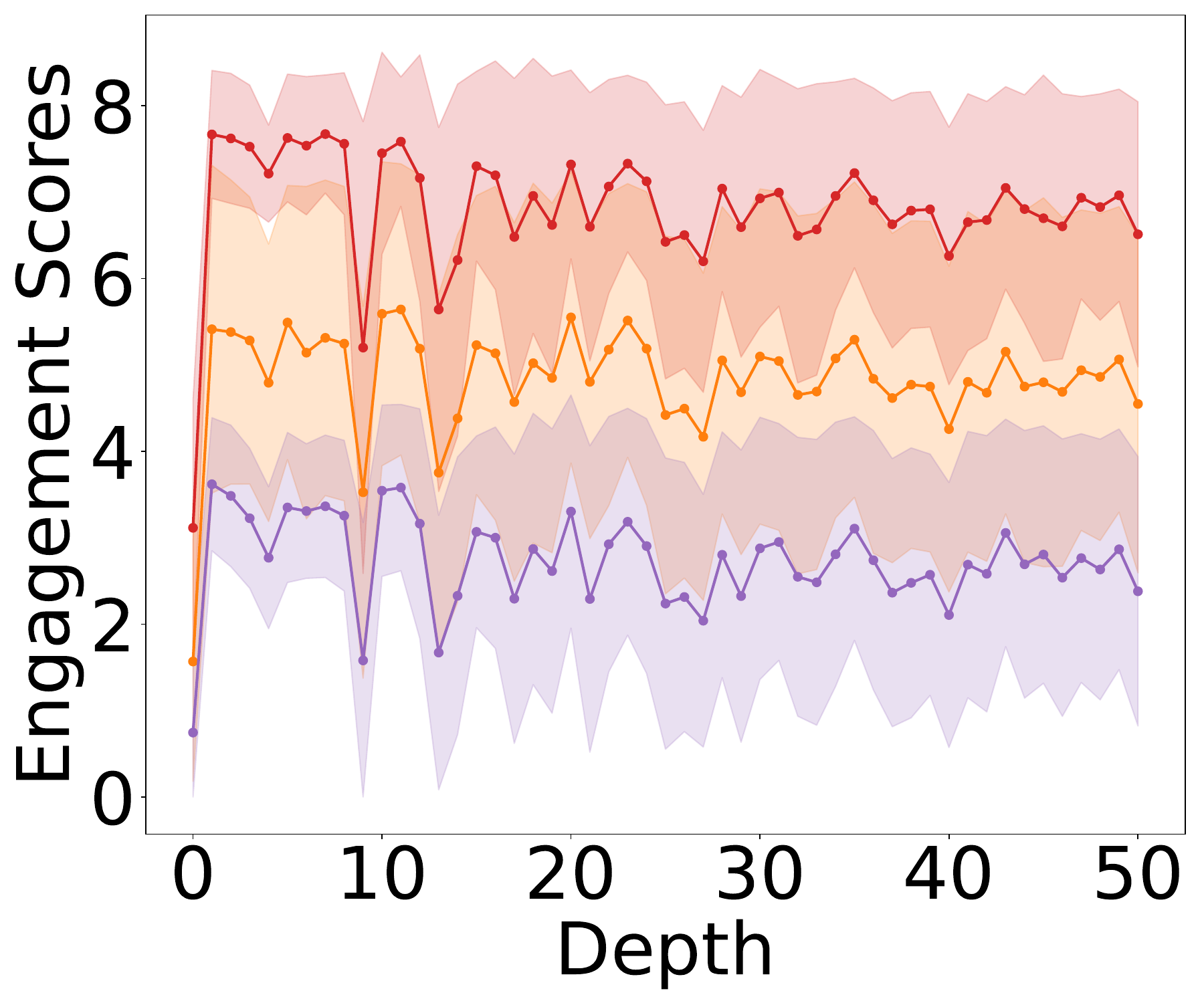}
    \caption{\centering \shortstack{Taiwan Engagements\\(60 Secs)}}
    \label{taiwan-engagement_60}
  \end{subfigure}

  \vspace{0cm}

  \begin{subfigure}[b]{0.32\textwidth}
    \includegraphics[width=\linewidth]{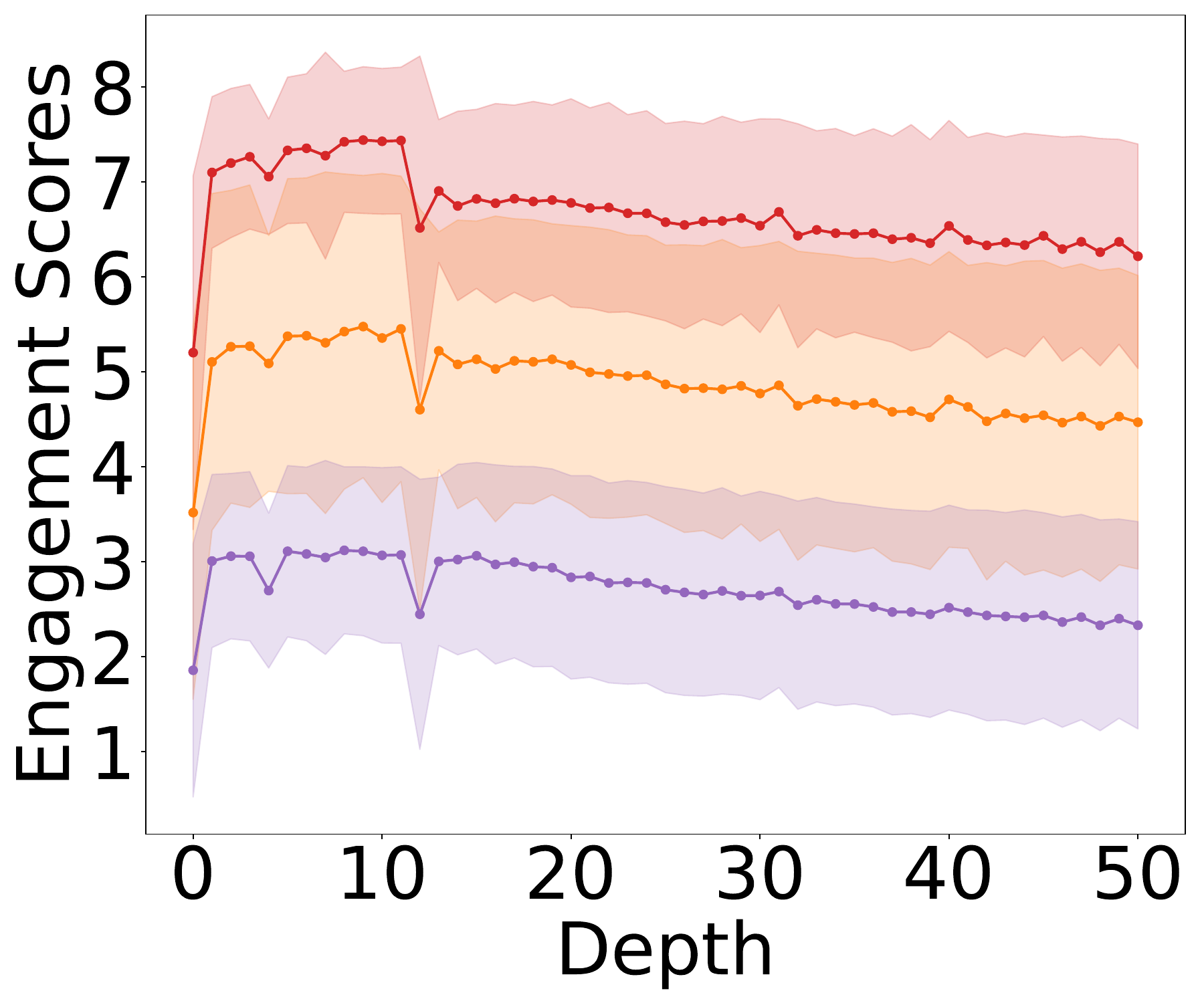}
    \caption{\centering \shortstack{General Engagements\\(3 Secs)}}
    \label{yt-general-engagement_3}
  \end{subfigure}
  \hfill
  \begin{subfigure}[b]{0.32\textwidth}
    \includegraphics[width=\linewidth]{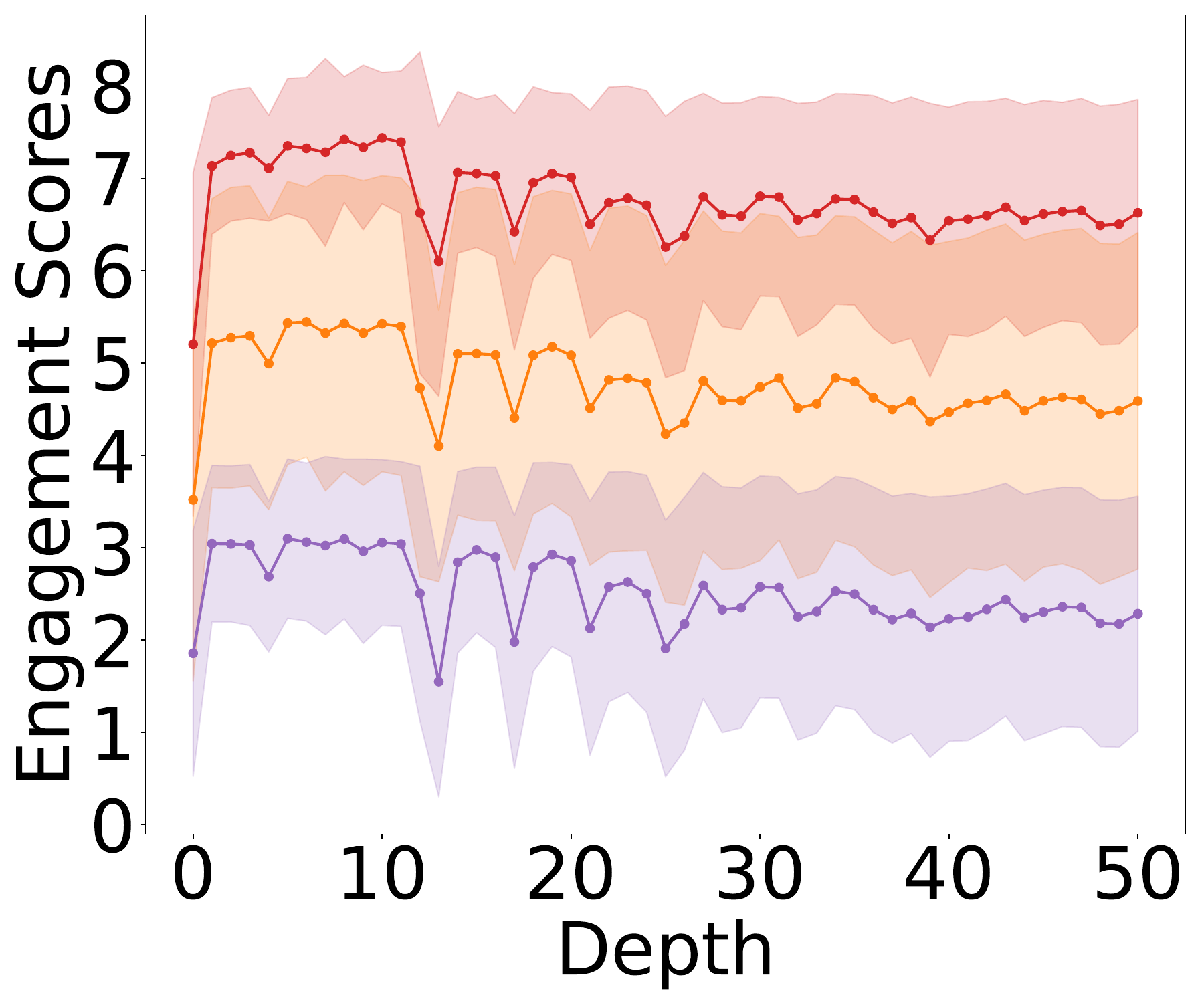}
    \caption{\centering \shortstack{General Engagements\\(15 Secs)}}
    \label{yt-general-engagement_15}
  \end{subfigure}
  \hfill
  \begin{subfigure}[b]{0.32\textwidth}
    \includegraphics[width=\linewidth]{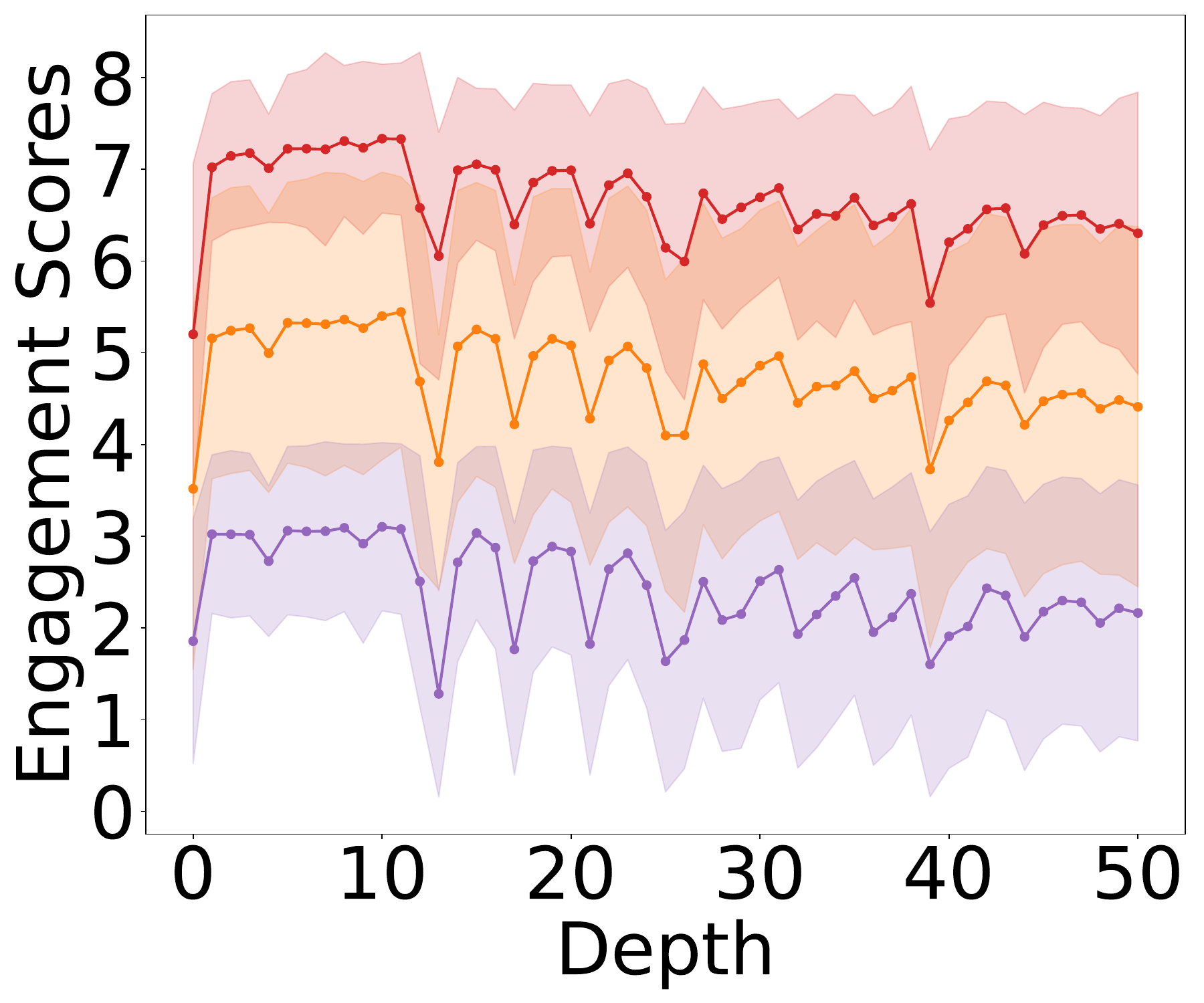}
    \caption{\centering \shortstack{General Engagements\\(60 Secs)}}
    \label{yt-general-engagement_60}
  \end{subfigure}

  \caption{Log-transformed engagement distribution across South China Sea (SCS), Taiwan Election, and general YouTube content videos at each recommendation depth. The different lines represent views, likes, and comments, showing the mean values at each depth.}

  \label{fig:all_engagament}
\end{figure}

To assess whether YouTube Shorts' recommendation algorithm favors high-engagement content, we analyzed three metrics: \textit{views}, \textit{likes}, and \textit{comments}, which serve as proxies for user interest. Due to scale differences—views often in millions, likes and comments much lower—we applied a base-10 logarithmic transformation to normalize the data. Engagement patterns are shown in Figure~\ref{fig:all_engagament}. Across all datasets and watch-time conditions, engagement scores rose sharply after the first recommendation depth, suggesting that the algorithm prioritizes videos with high interaction levels. Following this initial rise, engagement generally stabilized, with variation depending on watch-time. In the 60-second condition (Figures~\ref{scs-engagement_60}, \ref{taiwan-engagement_60}, \ref{yt-general-engagement_60}), we observed periodic fluctuations beyond depth 10, often coinciding with the appearance of ads or sponsored videos. These promotional insertions may be strategically timed once user attention is sustained, leading to temporary dips in engagement as the algorithm transitions between organic and monetized content. Overall, the findings indicate that YouTube Shorts favors popular, high-engagement videos. While effective for maximizing platform retention, this bias can marginalize niche or serious topics, especially politically sensitive content, reinforcing broader patterns of content drift observed in prior analyses.

\section{Conclusion and Discussion} \label{conclusion}

This study investigated algorithmic drift and bias in YouTube Shorts by analyzing how its recommendation system evolves across dimensions of topical relevance, content category, and emotional tone. We focused on two politically sensitive topics—South China Sea and Taiwan election—and compared them with general YouTube content. Using validated generative AI models and simulated watch-time conditions, we assessed how recommendations shift with depth and engagement. For \textbf{RQ1}, we observed a sharp drop in topical relevance immediately after the first recommendation for political content. This was accompanied by a clear drift toward unrelated entertainment or non-political videos, indicating a loss in topical diversity and coherence. In relation to \textbf{RQ2}, we found that entertainment content was consistently favored, while political topics were quickly deprioritized. Emotion analysis further revealed that joyful and neutral tones were promoted, whereas anger and fear diminished across depth. This suggests a platform-level preference for emotionally positive, engagement-optimized content. Addressing \textbf{RQ3}, we found that longer watch durations led to more sustained drift. Promoted or sponsored content beyond depth 10 was prominently periodic in longer watch-time conditions. Higher engagement also did not prevent drifting from the original topic. These results indicate that the watch time does not reduce algorithmic bias. In conclusion, YouTube Shorts' recommendation algorithm demonstrates a consistent shift away from politically sensitive or emotionally negative content, favoring high-engagement and emotionally positive videos. These findings raise broader concerns about content visibility, personalization dynamics, and the transparency of algorithmic curation in short-form video platforms. Future work will expand beyond political topics to include a broader range of themes, larger datasets, and other platforms. We also plan to compare politically engaged and neutral user profiles to assess the impact of personalization on algorithmic drift.

\bmhead{Acknowledgements}
\small
This research is funded in part by the U.S. NSF (OIA-1946391, OIA-1920920), OUSD/AFOSR (FA9550-22-1-0332), ARO (W911NF-23-1-0011, W911NF-24-1-0078), U.S. ONR (N00014-21-1-2121, N00014-21-1-2765, N00014-22-1-2318), AFRL, DARPA,  Australian Department of Defense Strategic Policy Grants Program, Arkansas Research Alliance, the Jerry L. Maulden/Entergy Endowment, and the Donaghey Foundation at the UA Little Rock. The researchers gratefully acknowledge the support.

\renewcommand{\bibfont}{\fontsize{8.5}{10.4}\selectfont}

\bibliography{sn-bibliography}

\end{document}